\documentclass[a4paper,10pt]{article}
\pdfoutput=1
\usepackage{tikz}

\usepackage{jcappub}
\usepackage{comment}

\usepackage[T1]{fontenc}
\usepackage{mathtools}
\usepackage{microtype}
\usepackage[capitalize]{cleveref}

\newcommand{\mpl}{m_\mathrm{Pl}}
\newcommand{\abs}[1]{{\left \vert #1 \right \vert}}
\newcommand{\overbar}[1]{\mkern 1.25mu\overline{\mkern-1.25mu#1\mkern-1.25mu}\mkern 1.25mu}

\newcommand{\ud}{\mathrm{d}}
\newcommand{\Beq}{\begin{align}}
\newcommand{\Eeq}{\end{align}}

\newcommand{\sunset}{Z}
\newcommand{\fourvertex}{C}

\title{Non-Gaussianity and the induced gravitational wave background}

\author[a]{Peter Adshead,}
\author[a]{Kaloian D. Lozanov,}
\author[a,b]{and Zachary J. Weiner}

\affiliation[a]{Illinois Center for Advanced Studies of the Universe \& Department of Physics, University of Illinois at Urbana-Champaign, Urbana, IL 61801, U.S.A.}
\affiliation[b]{Department of Physics, University of Washington, Seattle, WA 98195, U.S.A.}

\emailAdd{adshead@illinois.edu}
\emailAdd{klozanov@illinois.edu}
\emailAdd{zweiner@uw.edu}

\abstract{
    Scalar metric fluctuations generically source a spectrum of gravitational waves at second order
    in perturbation theory, poising gravitational wave experiments as potentially powerful probes
    of the small-scale curvature power spectrum.
    We perform a detailed study of the imprint of primordial non-Gaussianity on these induced
    gravitational waves, emphasizing the role of both the disconnected and connected components of
    the primoridal trispectrum.
    Specializing to local-type non-Gaussianity, we numerically compute all contributions and present
    results for a variety of enhanced primordial curvature power spectra.
}

\begin{document}
\maketitle
\flushbottom

\section{Introduction}

The paradigm of cosmic inflation~\cite{1979JETPL..30..682S,PhysRevD.23.347,Sato:1980yn,Mukhanov:1981xt,LINDE1982389,PhysRevLett.48.1220,1983PhLB..129..177L},
together with the $\Lambda$ cold dark matter ($\Lambda$CDM) model, provides a precise description of
the Universe on cosmic scales~\cite{Aghanim:2018eyx,Alam:2016hwk,Scolnic:2017caz,Akrami:2018odb}.
This model, in which the Universe expands from a hot and dense state following inflation,
successfully predicts both the primordial elemental abundances via big bang nucleosynthesis
(BBN)~\cite{Steigman:2007xt,Cyburt:2015mya} as well as the cosmic microwave background radiation
(CMB).
Furthermore, measurements of the temperature anisotropies of the CMB reveal a red-tilted spectrum of
adiabatic and highly Gaussian density fluctuations in excellent agreement with the predictions of
standard slow roll inflation~\cite{Akrami:2018odb}.

The large ($10-10^4$ Mpc) scales measured in the CMB and large-scale structure directly probe (and
constrain) the dynamics of inflation around $50-60$ $e$-folds before its end~\cite{Kite:2020uix}.
The later stages of inflation are only weakly constrained by (the nonobservation of) spectral
distortions in the CMB~\cite{Chluba:2015bqa}, as well as the absence of high-energy $\gamma$ rays
from ultracompact minihalos~\cite{Bringmann:2011ut}.
Moreover, the abundances of light elements are insensitive to the state of the Universe prior to BBN
and neutrino decoupling (before redshift $z \sim 10^{10}$)~\cite{Steigman:2007xt}.
In fact, successful BBN requires only that the Universe was in local thermal equilibrium and
expanding in a radiation dominated state by a temperature of around $T \sim 4.1$ MeV
\cite{Ichikawa:2005vw,deSalas:2015glj,Hasegawa:2019jsa}.

Gravitational waves (GW) offer a unique means to study inhomogeneities on smaller scales and thereby
constrain both the later stages of inflation~\cite{Inomata:2018epa} and the evolution of the
Universe before BBN~\cite{Domenech:2019quo,Inomata:2019zqy,Inomata:2019ivs,Hajkarim:2019nbx,Domenech:2020kqm,Allahverdi:2020bys}.
Aside from direct production mechanisms, density perturbations act as a secondary source of
stochastic GW backgrounds~\cite{10.1143/PTP.37.831,PhysRevD.47.1311,Matarrese:1993zf,Matarrese:1997ay,Carbone:2004iv,Ananda:2006af,Baumann:2007zm,Saito:2008jc}.
Future experiments~\cite{2017arXiv170200786A,Seto:2001qf,Yagi:2011wg,Maggiore:2019uih,Reitze:2019iox,Lentati:2015qwp,Bian:2020bps,Aggarwal:2020olq}
will probe these ``induced'' GWs at a variety of frequencies~\cite{Bugaev:2009zh,Bugaev:2010bb,Alabidi:2012ex,Alabidi:2013lya,Garcia-Bellido:2016dkw,Inomata:2016rbd,Orlofsky:2016vbd,Nakama:2016gzw,Garcia-Bellido:2017aan,Domenech:2017ems,Kohri:2018awv,Clesse:2018ogk,Inomata:2019zqy,Inomata:2019ivs,Chen:2019xse,Domenech:2019quo,Ota:2020vfn,Cai:2019jah,Yuan:2019wwo,Cai:2019elf,Cai:2019amo,Bartolo:2019zvb,Bhattacharya:2019bvk,Ozsoy:2019lyy,Pi:2020otn,Xu:2019bdp,Yuan:2020iwf,Ballesteros:2020qam,Liu:2020oqe,Braglia:2020eai,Braglia:2020taf,Fumagalli:2020nvq,Gow:2020bzo,Riccardi:2021rlf,Atal:2021jyo},
providing valuable information about scales that exit the horizon during inflation long after the
modes that eventually seed the CMB anisotropies.
These modes reenter the horizon before BBN and subsequently induce gravitational waves at second
order in cosmological perturbation theory.
Enticingly, the pulsar timing array experiment NANOGrav recently presented evidence for a common
process~\cite{Arzoumanian:2020vkk} which, though currently lacking Bayesian evidence for the
requisite quadrupolar correlations, may well be due to a stochastic GW
background~\cite{Vaskonen:2020lbd,DeLuca:2020agl,Kohri:2020qqd,Vagnozzi:2020gtf,Domenech:2020ers,Bhattacharya:2020lhc,Inomata:2020xad,Atal:2020yic,Pandey:2020gjy}.

The expected level of GWs induced from a red-tilted spectrum of curvature perturbations extrapolated
to small scales, with amplitude $\Delta_\mathcal{R}^2(k_\star) \approx 10^{-10}$ at $k_\star = 0.05
\, h$ Mpc$^{-1}$ required by the CMB anisotropies, is unobservably tiny
\cite{Ananda:2006af,Baumann:2007zm}.
However, there is no a priori reason to expect that such an extrapolation is appropriate over such
a large range of scales.
In particular, induced GWs are expected to be significant in scenarios where primordial black holes
(PBHs) form via the gravitational collapse of small-scale curvature
perturbations~\cite{Espinosa:2018eve,Cai:2018dig,Bartolo:2018rku,Unal:2018yaa,Yuan:2019udt}.
GWs therefore provide a powerful probe not only of the initial conditions and
expansion history of the Universe but also of the abundance of PBHs and their potential to
constitute a sizable fraction of the dark matter~\cite{Carr:2016drx,Inomata:2017okj,Green:2020jor}.
For recent reviews, see Refs.~\cite{Carr:2020gox,Carr:2020xqk,Sasaki:2018dmp,Yuan:2021qgz}.

While the running of the spectral index typically suppresses power on very small scales in canonical
models of inflation (see, e.g.~\cite{Adshead:2010mc}), a number of scenarios can enhance the
curvature power spectrum on small scales.
Examples of modifications to the minimal slow roll scenario include multifield
inflation~\cite{Frampton:2010sw,Kawasaki:2012kn}, inflaton couplings leading to particle
production~\cite{Anber:2009ua,Linde:2012bt,Bugaev:2013fya,Garcia-Bellido:2016dkw,Domcke:2017fix,Garcia-Bellido:2017aan,Garcia:2020mwi,Ozsoy:2020ccy,Ozsoy:2020kat},
a plateau in the inflaton potential that causes a period of ultra slow
roll~\cite{PhysRevD.50.7173,Byrnes:2018txb,Ozsoy:2018flq,Ragavendra:2020sop}, and a brief downward step in the
potential~\cite{Inomata:2021uqj}.
In these scenarios the curvature perturbation is often non-Gaussian on these scales, impacting not
only PBH formation (which is sensitive to the tail of the probability distribution of density
perturbations) but also the induced GW signal.

In this work we revisit the problem of GW backgrounds induced by non-Gaussian curvature
perturbations~\cite{Nakama:2016gzw,Garcia-Bellido:2017aan,Unal:2018yaa,Cai:2018dig,Cai:2019amo,Yuan:2020iwf,Ragavendra:2020sop,Atal:2021jyo}.
We focus on local-type non-Gaussianity~\cite{Unal:2018yaa,Cai:2018dig,Cai:2019amo,Yuan:2020iwf,Ragavendra:2020sop,Atal:2021jyo}
and show that the induced GW spectrum is in general sensitive to all contributions to the primordial
trispectrum.
In particular, in addition to the disconnected part of the curvature perturbation's 4-point
correlation function (arising solely via the non-Gaussian modification to the curvature power
spectrum itself), the connected component is a nontrivial and important contribution to the induced
GW signal, often unaccounted for in the literature.

This paper is organized as follows.
In \cref{sec:background} we briefly review the calculation of the dimensionless power spectrum
$\mathcal{P}_\lambda$ of GWs induced by a generic curvature perturbation $\mathcal{R}$, leaving
additional details to \cref{app:cpt}.
We proceed to present the complete contribution of the (connected and disconnected parts of the)
4-point correlation function of $\mathcal{R}$ to $\mathcal{P}_\lambda$.
We then specialize to the case of local non-Gaussianity.
\Cref{app:trispectrum} outlines the derivation in more detail, presenting a diagrammatic
interpretation of individual terms.
We present results for a variety of primordial curvature power spectra in \cref{sec:results} and
conclude in \cref{sec:conclusion}.

\section{Gravitational waves induced by scalar perturbations}\label{sec:background}

We work with a perturbed FLRW spacetime in the conformal Newtonian gauge,
\begin{align}
    \ud s^2
    &= a(\tau)^2 \left(
            - \left[ 1 + 2 \Phi \right] \ud \tau^2
            + \left[ (1 - 2 \Phi) \delta_{ij} + \frac{1}{2} h_{ij} \right]
            \ud x^i \ud x^j
        \right),
\end{align}
neglecting vector perturbations and first-order tensor perturbations.
That is, we consider only tensors sourced at second order in perturbation theory and ignore those at
leading order, e.g., a primordial background from inflation.
We set $c = \hbar = k_B = 1$, define the reduced Planck mass $\mpl = 1 / \sqrt{8 \pi G_N}$, and use
primes to denote derivatives with respect to conformal time $\tau$.
Repeated Latin indices denote a contraction with the Kronecker delta regardless of placement.
We assume a fixed equation of state $w = \overbar{P} / \overbar{\rho}$, with $\overbar{P}$ and
$\overbar{\rho}$ the background pressure and energy density, so the scale factor evolves according
to
\begin{align}\label{eqn:scale-factor-fixed-eos}
    a(\tau)
    &= a_0 \left( \frac{\tau}{\tau_0} \right)^\alpha,
\end{align}
where $\alpha = 2 / (1 + 3 w)$.
The conformal-time Hubble parameter is thus
\begin{align}
    \mathcal{H}(\tau)
    \equiv \frac{a'(\tau)}{a(\tau)}
    &= \frac{\alpha}{\tau}.
\end{align}

We expand the GWs in Fourier modes as
\begin{align}
    h_{ij}(\tau,\mathbf{x})
    &= \sum_{\lambda=+,\times}
        \int \frac{\ud^3 k}{(2\pi)^{3/2}} e^{i\mathbf{k}\cdot\mathbf{x}}
        \epsilon_{ij}^{\lambda}(\mathbf{k}) h_\lambda(\tau, \mathbf{k}),
\end{align}
where the polarization tensors are
\begin{subequations}\label{eqn:q-basis-def}
\begin{align}
    \epsilon^+_{ij}(\mathbf{k})
    &= \frac{1}{\sqrt{2}} \left(
        \epsilon_i(\mathbf{k}) \epsilon_j(\mathbf{k})
        - \overbar{\epsilon}_i(\mathbf{k}) \overbar{\epsilon}_j(\mathbf{k})
     \right) \\
    \epsilon^\times_{ij}(\mathbf{k})
    &= \frac{1}{\sqrt{2}} \left(
            \epsilon_i(\mathbf{k}) \overbar{\epsilon}_j(\mathbf{k})
            + \overbar{\epsilon}_i(\mathbf{k}) \epsilon_j(\mathbf{k})
        \right)
\end{align}
\end{subequations}
and $\epsilon_{i}(\mathbf{k})$ and $\overbar{\epsilon}_{i}(\mathbf{k})$ form an orthonormal basis
transverse to $\mathbf{k}$.
Note that $\epsilon_{ij}^{\lambda}(\mathbf{k})$ are traceless and transverse to $\mathbf{k}$ by
construction.
The power spectrum of GWs is then defined as
\begin{align}
    \langle
        h_{\lambda_1}(\tau, \mathbf{k}_1)
        h_{\lambda_2}(\tau, \mathbf{k}_2)
    \rangle
    &= \delta^3(\mathbf{k}_1 + \mathbf{k}_2)
        \delta^{\lambda_1 \lambda_2}
        \mathcal{P}_{\lambda_1}(\tau, k_1).
\end{align}
We also define the dimensionless GW power spectrum,
\begin{align}
    \langle
        h_{\lambda_1}(\tau, \mathbf{k}_1)
        h_{\lambda_2}(\tau, \mathbf{k}_2)
    \rangle
    &= \delta^3(\mathbf{k}_1 + \mathbf{k}_2)
        \delta^{\lambda_1 \lambda_2}
        \frac{2\pi^2}{k_1^3}
        \Delta^2_{\lambda_1}(\tau, k_1).
\end{align}

The spatially averaged energy density of gravitational waves on subhorizon scales is
\begin{align}
    \rho_\mathrm{GW}(\tau)
    &= \int \ud \ln k \, \rho_\mathrm{GW} (\tau, k)
    = \frac{\mpl^2}{16 a(\tau)^2}
    \left\langle \overbar{\partial_k h_{ij} \partial_k h_{ij}} \right\rangle,
\end{align}
where the overbar denotes a time average (i.e., over oscillations).
The fractional energy density in GWs per logarithmic wavenumber is
\begin{align}
    \Omega_\mathrm{GW}(\tau, k)
    \equiv \frac{\rho_\mathrm{GW}(\tau, k)}{\rho_\mathrm{tot}(\tau)}
    = \frac{1}{48}
        \left(\frac{k}{a(\tau)H(\tau)}\right)^2
        \sum_{\lambda=+,\times} \overbar{\Delta^2_\lambda(\tau,k)}.
\end{align}
The spectrum that would be observed today (assuming emission after reheating) is obtained via the
transfer function
\begin{align}
    \Omega_{\mathrm{GW}, 0}(k) h^2
    &= \Omega_{\mathrm{rad}, 0} h^2
        \left( \frac{g_{\star, 0}}{g_{\star, \mathrm{e}}} \right)^{1/3}
        \Omega_{\mathrm{GW}, \mathrm{e}}(k),
    \label{eqn:gw-amplitude-transfer-function}
\end{align}
which for comoving momentum $k$ would be observed at the present-day frequency
\begin{align}
    f
    = \frac{k / 2 \pi a_\mathrm{e}}{\sqrt{H_\mathrm{e} \mpl}}
        \left(
            \Omega_{\mathrm{rad}, 0}
            H_0^2 \mpl^2
        \right)^{1/4}
        \left( \frac{g_{\star, 0}}{g_{\star, \mathrm{e}}} \right)^{1/12}.
\end{align}
Here $g_{\star}$ is the number of relativistic degrees of freedom in energy
density, $\Omega_{\mathrm{rad}, 0}$ the present-day abundance of radiation, and $H$ the Hubble
parameter; subscripts $\mathrm{e}$ and $0$ denote the time of emission and the present day,
respectively.
Note that $\Omega_{\mathrm{rad}, 0} h^2 \approx 4.2 \times 10^{-5}$ with
$h = H_0 / 100 \, \mathrm{km} \mathrm{s}^{-1} / \mathrm{Mpc}$.

\subsection{Induced gravitational wave solution}
\label{sec:GWdyn}

The induced gravitational waves evolve according to
\begin{align}
\label{eq:GWeom}
    h_\lambda''(\tau, \mathbf{k})
        + 2 \mathcal{H} h_\lambda'(\tau, \mathbf{k})
        + k^2 h_\lambda(\tau, \mathbf{k})
    &= 4 \mathcal{S}_\lambda(\tau, \mathbf{k}),
\end{align}
where the source $\mathcal{S}_\lambda$ comprises the  terms of the (transverse, traceless part of
the) Einstein equation  that are second order in scalar perturbations.
To make contact with primordial physics, it is conventional to express the gravitational potential
in terms of the primordial curvature perturbation $\mathcal{R}$ and a transfer function $\Phi(k\tau)$,
\begin{align}\label{eqn:transfer-times-curvature}
    \Phi(\tau, \mathbf{k})
    &= \frac{3+3w}{5+3w} \Phi(k \tau) \mathcal{R}(\mathbf{k}).
\end{align}
The transfer function $\Phi(k\tau)$ encodes the linear evolution of the Newtonian potential after
horizon reentry.
In these terms, the source is~\cite{Ananda:2006af,Baumann:2007zm}
\begin{align}
    \label{eq:S}
    \mathcal{S}_\lambda(\tau, \mathbf{k})
    &= \int \frac{\ud^3 q}{(2\pi)^{3/2}} Q_{\lambda}(\mathbf{k},\mathbf{q})
        f(\vert \mathbf{k}-\mathbf{q} \vert, q, \tau)
        \mathcal{R}(\mathbf{q})
        \mathcal{R}(\mathbf{k}-\mathbf{q}),
\end{align}
where $f(\vert \mathbf{k}-\mathbf{q} \vert, q, \tau)$ is given by
\begin{align}\label{eqn:source-function-p-q-tau}
\begin{split}
    f(p, q, \tau)
    &= \frac{3 (1 + w)}{(5 + 3 w)^2}
        \Big[
            2 (5 + 3 w)
            \Phi(p \tau) \Phi(q \tau)
            + \tau^2 \left( 1 + 3 w \right)^2 \Phi'(p \tau) \Phi'(q \tau)
    \\ &\hphantom{ {}={}
        \frac{3 (1 + w)}{(5 + 3 w)^2}
        \Big[
    }
            + 2 \tau \left( 1 + 3 w \right)
            \left(
                \Phi(p \tau) \Phi'(q \tau) + \Phi'(p \tau) \Phi(q \tau)
            \right)
        \Big].
\end{split}
\end{align}
Note that $f(p, q, \tau)$ is symmetric under exchange of $q$ and $p$.
The projection factors are
\begin{align}
    \label{eqn:projection-factor-def}
    Q_{\lambda}(\mathbf{k}, \mathbf{q})
    &\equiv \epsilon_{ij}^{\lambda}(\mathbf{k}) q_i q_j.
\end{align}
Taking $\mathbf{k}$ in the $\hat{z}$ direction and writing
\begin{align}
    \mathbf{q}
    &= q
        \left[
            \sin \theta \cos \phi,
            \sin \theta \sin \phi,
            \cos \theta
        \right],
\end{align}
the projection factors evaluate to
\begin{align}\label{eqn:projection-factor-ito-angles}
    Q_{\lambda}(\mathbf{k},\mathbf{q})
    &=
    \frac{q^2}{\sqrt{2}}
    \sin^2(\theta)
    \times
    \begin{cases}
        \cos(2\phi) &\lambda = + \\
        \sin(2\phi) &\lambda = \times.
    \end{cases}
\end{align}
We note that the $\cos(2\phi)$ and $\sin(2\phi)$ terms are absent in some of the intermediate steps
in Ref.~\cite{Baumann:2007zm}.

The induced GWs are the particular solution of \cref{eq:GWeom},
\begin{align}
\label{eq:h}
    h_\lambda(\tau, \mathbf{k})
    = \frac{4}{a(\tau)} \int^{\tau}_{\tau_0} \ud \overbar{\tau} \,
        G_\mathbf{k}(\tau, \overbar{\tau})
        a(\overbar{\tau})
        \mathcal{S}_\lambda(\overbar{\tau}, \mathbf{k}),
\end{align}
where the Green function $G_\mathbf{k }(\tau,\overbar{\tau})$ obeys the equation of motion
\begin{align}\label{eqn:green-function-eom}
    \partial_{\tau}^2 G_\mathbf{k}(\tau,\overbar{\tau})
        + \left[ k^2 - \frac{a''(\tau)}{a(\tau)} \right]
        G_\mathbf{k}(\tau, \overbar{\tau})
    &= \delta(\tau - \overbar{\tau}).
\end{align}
Hence, the power spectrum of the induced GWs is given by
\begin{align}
\begin{split}\label{eqn:gw-power-spectrum-generic}
    \langle
        h_{\mathbf{k}_1}^{\lambda_1}
        h_{\mathbf{k}_2}^{\lambda_2}
    \rangle
    &= 16
        \int \frac{\ud^3 q_1}{(2\pi)^{3/2}} \frac{\ud^3 q_2}{(2\pi)^{3/2}}
        \langle
            \mathcal{R}(\mathbf{q}_1)
            \mathcal{R}(\mathbf{k}_1 - \mathbf{q}_1)
            \mathcal{R}(\mathbf{q}_2)
            \mathcal{R}(\mathbf{k}_2 - \mathbf{q}_2)
        \rangle
        Q_{\lambda_1}(\mathbf{k}_1, \mathbf{q}_1)
        Q_{\lambda_2}(\mathbf{k}_2, \mathbf{q}_2) \\
    &\hphantom{{}={} 16
            \int \frac{\ud^3 q_1}{(2\pi)^{3/2}} \frac{\ud^3 q_2}{(2\pi)^{3/2}}
        }
        \times
        I(\abs{ \mathbf{k}_1 - \mathbf{q}_1 }, q_1, \tau_1)
        I(\abs{ \mathbf{k}_2 - \mathbf{q}_2 }, q_2, \tau_2),
\end{split}
\end{align}
having defined
\begin{align}\label{eqn:I-def}
    I(p, q, \tau)
    &= \int_{\tau_0}^\tau \ud \overbar{\tau} \,
        G_{\mathbf{k}}(\tau, \overbar{\tau}) \frac{a(\overbar{\tau})}{a(\tau)}
        f(p, q, \overbar{\tau}).
\end{align}
By treating the scalar perturbations to linear order, the time dependence of the induced GW spectrum
is decoupled from the primordial curvature power spectrum.
For a fixed equation of state, the integrals $I(p, q, \tau)$ can be computed
analytically~\cite{Kohri:2018awv}.

If we impose statistical homogeneity and isotropy on the curvature perturbation $\mathcal{R}$ and
assume $\langle \mathcal{R} \rangle = 0$, then its 4-point function splits into disconnected and
connected components:
\begin{align}
    \langle
        \mathcal{R}(\mathbf{k}_1)
        \mathcal{R}(\mathbf{k}_2)
        \mathcal{R}(\mathbf{k}_3)
        \mathcal{R}(\mathbf{k}_4)
    \rangle
    &= \langle
            \mathcal{R}(\mathbf{k}_1)
            \mathcal{R}(\mathbf{k}_2)
            \mathcal{R}(\mathbf{k}_3)
            \mathcal{R}(\mathbf{k}_4)
        \rangle_\mathrm{d}
        +\langle
            \mathcal{R}(\mathbf{k}_1)
            \mathcal{R}(\mathbf{k}_2)
            \mathcal{R}(\mathbf{k}_3)
            \mathcal{R}(\mathbf{k}_4)
        \rangle_\mathrm{c} \\
\begin{split}
    \langle
        \mathcal{R}(\mathbf{k}_1)
        \mathcal{R}(\mathbf{k}_2)
        \mathcal{R}(\mathbf{k}_3)
        \mathcal{R}(\mathbf{k}_4)
    \rangle_\mathrm{d}
    &= \langle
            \mathcal{R}(\mathbf{k}_1)
            \mathcal{R}(\mathbf{k}_2)
        \rangle
        \langle
            \mathcal{R}(\mathbf{k}_3)
            \mathcal{R}(\mathbf{k}_4)
        \rangle
        + \langle
            \mathcal{R}(\mathbf{k}_2)
            \mathcal{R}(\mathbf{k}_3)
        \rangle
        \langle
            \mathcal{R}(\mathbf{k}_4)
            \mathcal{R}(\mathbf{k}_1)
        \rangle
    \\ &\hphantom{ {}={} }
        + \langle
            \mathcal{R}(\mathbf{k}_1)
            \mathcal{R}(\mathbf{k}_3)
        \rangle
        \langle
            \mathcal{R}(\mathbf{k}_2)
            \mathcal{R}(\mathbf{k}_4)
        \rangle
\end{split} \\
    \langle
        \mathcal{R}(\mathbf{k}_1)
        \mathcal{R}(\mathbf{k}_2)
        \mathcal{R}(\mathbf{k}_3)
        \mathcal{R}(\mathbf{k}_4)
    \rangle_\mathrm{c}
    &=
        \delta^3(\mathbf{k}_1 + \mathbf{k}_2 + \mathbf{k}_3 + \mathbf{k}_4)
        \mathcal{T}(\mathbf{k}_1, \mathbf{k}_2, \mathbf{k}_3, \mathbf{k}_4),
\end{align}
where
\begin{align}
    \langle
        \mathcal{R}(\mathbf{k}_1)
        \mathcal{R}(\mathbf{k}_2)
    \rangle
    = \delta^3(\mathbf{k}_1 + \mathbf{k}_2)
        \mathcal{P}_{\mathcal{R}}(k_1),
\end{align}
defines the power spectrum of the curvature perturbation and
$\mathcal{T}(\mathbf{k}_1,\mathbf{k}_2,\mathbf{k}_3,\mathbf{k}_4)$ stands for the connected
trispectrum.
We similarly decompose the induced GW power spectrum into its parts sourced by the disconnected and
connected trispectrum:\footnote{
    Note that nonlinear terms at higher order (in $\Phi$) in the source term, \cref{eq:S}, are
    neglected here.
    Provided $F_\mathrm{NL} > \mathcal{O}(1)$ (see \cref{eq:RNG} below), the contribution from
    primordial non-Gaussianity dominates.
}
\begin{align}
    \mathcal{P}_\lambda(k)
    &= \left. \mathcal{P}_\lambda(k) \right\vert_\mathrm{d}
        + \left. \mathcal{P}_\lambda(k) \right\vert_\mathrm{c}.
\end{align}
The disconnected component represents that arising from the curvature power spectrum, including its
contribution from non-Gaussianity (the exact form of which we have yet to specify).
This term takes the form
\begin{align}\label{eqn:P-h-disc-generic}
    \left. \mathcal{P}_\lambda(k) \right\vert_\mathrm{d}
    &= 32
        \int \frac{\ud^3 q}{(2\pi)^3}
        Q_{\lambda}(\mathbf{k}, \mathbf{q})^2
        I(\abs{ \mathbf{k} - \mathbf{q} }, q, \tau)^2
        \mathcal{P}_\mathcal{R}(q)
        \mathcal{P}_\mathcal{R}(\abs{\mathbf{k} - \mathbf{q}}).
\end{align}
In the case that $\mathcal{R}$ is a Gaussian field, \cref{eqn:P-h-disc-generic} reproduces the
standard result~\cite{Ananda:2006af,Baumann:2007zm,Kohri:2018awv}.
The connected component is
\begin{align}
\begin{split}\label{eqn:P-h-conn-generic}
    \left. \mathcal{P}_\lambda(k) \right\vert_\mathrm{c}
    &= 16
        \int \frac{\ud^3 q_1}{(2\pi)^{3/2}} \frac{\ud^3 q_2}{(2\pi)^{3/2}}
        Q_{\lambda}(\mathbf{k}, \mathbf{q}_1)
        Q_{\lambda}(\mathbf{k}, \mathbf{q}_2)
        I(\abs{ \mathbf{k} - \mathbf{q}_1 }, q_1, \tau)
        I(\abs{ \mathbf{k} - \mathbf{q}_2 }, q_2, \tau)
    \\ &\hphantom{{}={} 16
            \int \frac{\ud^3 q_1}{(2\pi)^{3/2}} \frac{\ud^3 q_2}{(2\pi)^{3/2}}
        }
        \times
        \mathcal{T}(
            \mathbf{q}_1, \mathbf{k} - \mathbf{q}_1,
            - \mathbf{q}_2, \mathbf{q}_2 - \mathbf{k}
        ).
\end{split}
\end{align}
We stress that the connected trispectrum contribution to the induced GWs power spectrum,
\cref{eqn:P-h-conn-generic}, does not vanish in general.
As we demonstrate explicitly below, the connected trispectrum generally has nontrivial dependence on
the azimuthal angles of $\mathbf{q}_1$ and $\mathbf{q}_2$.
Were this not the case, the azimuthal dependence would arise solely via
\cref{eqn:projection-factor-ito-angles}, and therefore the integrals over $\phi_1$ and $\phi_2$
would each vanish.\footnote{
    In Refs.~\cite{Cai:2018dig,Cai:2019amo,Ragavendra:2020sop},
    the contribution of the connected part of the trispectrum is neglected.
    Refs.~\cite{Cai:2018dig, Cai:2019amo} claim that the contribution from the connected terms
    vanishes due to the integrals over the azimuthal angles, regardless of the form of the
    trispectrum.}
Finally, extracting the observable GW signal requires taking the time average of the right-hand
side; see \cref{app:cpt} for details.

\subsection{Local-type non-Gaussian curvature as a source of GWs}
\label{sec:NGGWdyn}

We now specialize to the case of local-type non-Gaussianity,
\begin{align}
    \label{eq:RNG}
    \mathcal{R}(\mathbf{x})
    &= \mathcal{R}_g(\mathbf{x})
        + F_\mathrm{NL}
        \left( \mathcal{R}_g(\mathbf{x})^2 - \langle \mathcal{R}_g(\mathbf{x})^2 \rangle \right),
\end{align}
where the Gaussian field $\mathcal{R}_g$ is completely specified by its power spectrum, defined by
\begin{align}
    \left\langle \mathcal{R}_g(\mathbf{k}_1) \mathcal{R}_g(\mathbf{k}_2) \right\rangle
    &=
        \delta^3(\mathbf{k}_1 + \mathbf{k}_2) \mathcal{P}_g(k_1).
\end{align}
The one-loop power spectrum of $\mathcal{R}$ is
\begin{align}
    \mathcal{P}_\mathcal{R}(k)
    &= \mathcal{P}_g(k)
        + 2 F_\mathrm{NL}^2
        \int \frac{\ud^3 q}{(2 \pi)^3}
        \mathcal{P}_g(q)
        \mathcal{P}_g(\abs{\mathbf{k} - \mathbf{q}}).
\end{align}
Thus, there are three unique, disconnected contributions to the induced GW spectrum: the standard
Gaussian term,
\begin{align}\label{eqn:P-lambda-gaussian}
    \mathcal{P}_\lambda(k)_\mathrm{Gaussian}
    &= 2^5
        \int \frac{\ud^3 q}{(2 \pi)^3}
        I(\abs{\mathbf{k} - \mathbf{q}}, q, \tau)^2
        Q_\lambda(\mathbf{k}, \mathbf{q})^2
        \mathcal{P}_g(q)
        \mathcal{P}_g(\abs{\mathbf{k} - \mathbf{q}}),
\end{align}
an $\mathcal{O}(F_\mathrm{NL}^2)$ ``hybrid'' term,
\begin{align}\label{eqn:P-lambda-hybrid}
    \mathcal{P}_\lambda(k)_\mathrm{hybrid}
    &= 2^7 F_\mathrm{NL}^2
        \int \frac{\ud^3 q_1}{(2 \pi)^3}
        \frac{\ud^3 q_2}{(2 \pi)^3}
        I(\abs{\mathbf{k} - \mathbf{q}_1}, q_1, \tau)^2
        Q_\lambda(\mathbf{k}, \mathbf{q}_1)^2
        \mathcal{P}_g(\abs{\mathbf{k} - \mathbf{q}_1})
        \mathcal{P}_g(q_2)
        \mathcal{P}_g(\abs{\mathbf{q}_1 - \mathbf{q}_2}),
\end{align}
and an $\mathcal{O}(F_\mathrm{NL}^4)$ ``reducible'' term,
\begin{align}
\begin{split}\label{eqn:P-lambda-reducible}
    \mathcal{P}_\lambda(k)_\mathrm{reducible}
    &= 2^7 F_\mathrm{NL}^4
        \int \frac{\ud^3 q_1}{(2 \pi)^3}
        \frac{\ud^3 q_2}{(2 \pi)^3}
        \frac{\ud^3 q_3}{(2 \pi)^3}
        I(\abs{\mathbf{k} - \mathbf{q}_1}, q_1, \tau)^2
        Q_\lambda(\mathbf{k}, \mathbf{q}_1)^2
    \\ &\hphantom{{}={} 2^7 F_\mathrm{NL}^2 \int}
        \times
        \mathcal{P}_g(q_2)
        \mathcal{P}_g(q_3)
        \mathcal{P}_g(\abs{\mathbf{q}_1 - \mathbf{q}_2})
        \mathcal{P}_g(\abs{\mathbf{k} - \mathbf{q}_1 - \mathbf{q}_3}).
\end{split}
\end{align}
The connected contributions comprise an $\mathcal{O}(F_\mathrm{NL}^2)$ ``\fourvertex{}'' term,
\begin{align}
\begin{split}\label{eqn:P-lambda-4-vertex}
    \mathcal{P}_\lambda(k)_\mathrm{\fourvertex{}}
    &= 2^8
        F_\mathrm{NL}^2
        \int \frac{\ud^3 q_1}{(2\pi)^3} \frac{\ud^3 q_2}{(2\pi)^3}
        I(\abs{\mathbf{k} - \mathbf{q}_1}, q_1, \tau)
        I(\abs{\mathbf{k} - \mathbf{q}_2}, q_2, \tau)
        Q_\lambda(\mathbf{k}, \mathbf{q}_1)
        Q_\lambda(\mathbf{k}, \mathbf{q}_2)
    \\ &\hphantom{{}={} 2^9 F_\mathrm{NL}^2 \int }
        \times
        \mathcal{P}_g(q_2)
        \mathcal{P}_g(\abs{\mathbf{k} - \mathbf{q}_2})
        \mathcal{P}_g(\abs{\mathbf{q}_1 - \mathbf{q}_2}),
\end{split}
\end{align}
an $\mathcal{O}(F_\mathrm{NL}^2)$ ``\sunset{}'' term,
\begin{align}
\begin{split}\label{eqn:P-lambda-sunset}
    \mathcal{P}_\lambda(k)_\mathrm{\sunset{}}
    &= 2^8
        F_\mathrm{NL}^2
        \int \frac{\ud^3 q_1}{(2\pi)^3} \frac{\ud^3 q_2}{(2\pi)^3}
        I(\abs{\mathbf{k} - \mathbf{q}_1}, q_1, \tau)
        I(\abs{\mathbf{k} - \mathbf{q}_2}, q_2, \tau)
        Q_\lambda(\mathbf{k}, \mathbf{q}_1)
        Q_\lambda(\mathbf{k}, \mathbf{q}_2)
    \\ &\hphantom{{}={} 2^9 F_\mathrm{NL}^2 \int }
        \times
        \mathcal{P}_g(q_1)
        \mathcal{P}_g(q_2)
        \mathcal{P}_g(\abs{\mathbf{k} - (\mathbf{q}_1 + \mathbf{q}_2)}),
\end{split}
\end{align}
an $\mathcal{O}(F_\mathrm{NL}^4)$ ``planar'' term,
\begin{align}
\begin{split}\label{eqn:P-lambda-planar}
    \mathcal{P}_\lambda(k)_\mathrm{planar}
    &= 2^9
        F_\mathrm{NL}^4
        \int
        \frac{\ud^3 q_1}{(2\pi)^3}
        \frac{\ud^3 q_2}{(2\pi)^3}
        \frac{\ud^3 q_3}{(2\pi)^3}
        I(\abs{\mathbf{k} - \mathbf{q}_1}, q_1, \tau)
        I(\abs{\mathbf{k} - \mathbf{q}_2}, q_2, \tau)
        Q_\lambda(\mathbf{k}, \mathbf{q}_1)
        Q_\lambda(\mathbf{k}, \mathbf{q}_2)
    \\ &\hphantom{{}={} 2^{10} F_\mathrm{NL}^4 \int }
        \times
        \mathcal{P}_g(q_3)
        \mathcal{P}_g(\abs{\mathbf{q}_1 - \mathbf{q}_3})
        \mathcal{P}_g(\abs{\mathbf{q}_2 - \mathbf{q}_3})
        \mathcal{P}_g(\abs{\mathbf{k} - \mathbf{q}_3}),
\end{split}
\end{align}
and an $\mathcal{O}(F_\mathrm{NL}^4)$ ``nonplanar'' term,
\begin{align}
\begin{split}\label{eqn:P-lambda-nonplanar}
    \mathcal{P}_\lambda(k)_\mathrm{nonplanar}
    &= 2^8
        F_\mathrm{NL}^4
        \int
        \frac{\ud^3 q_1}{(2\pi)^3}
        \frac{\ud^3 q_2}{(2\pi)^3}
        \frac{\ud^3 q_3}{(2\pi)^3}
        I(\abs{\mathbf{k} - \mathbf{q}_1}, q_1, \tau)
        I(\abs{\mathbf{k} - \mathbf{q}_2}, q_2, \tau)
        Q_\lambda(\mathbf{k}, \mathbf{q}_1)
        Q_\lambda(\mathbf{k}, \mathbf{q}_2)
    \\ &\hphantom{{}={} 2^9 F_\mathrm{NL}^4 \int }
        \times
        \mathcal{P}_g(q_3)
        \mathcal{P}_g(\abs{\mathbf{q}_1 - \mathbf{q}_3})
        \mathcal{P}_g(\abs{\mathbf{q}_2 - \mathbf{q}_3})
        \mathcal{P}_g(\abs{\mathbf{k} - (\mathbf{q}_1 + \mathbf{q}_2) + \mathbf{q}_3}).
\end{split}
\end{align}
Ref.~\cite{Atal:2021jyo} refers to $\mathcal{P}_\lambda(k)_\mathrm{\sunset{}}$ as having a ``walnut''
topology; in \cref{app:trispectrum} we provide a complete prescription for assigning Feynman-type
diagrams to the above integrals (from which we derive the labels Z and C).
The ``walnut'' integral of Ref.~\cite{Unal:2018yaa} appears to be our
$\mathcal{P}_\lambda(k)_\mathrm{\fourvertex{}}$.
Furthermore, $\mathcal{P}_\lambda(k)_\mathrm{\fourvertex{}}$ does not appear in
Ref.~\cite{Atal:2021jyo}, while $\mathcal{P}_\lambda(k)_\mathrm{\sunset{}}$ does not appear in
Ref.~\cite{Unal:2018yaa}.
In \cref{app:recasting-the-integrals} we recast these integrals into a form suitable for
numerical integration.

\section{Results}\label{sec:results}

To study the relative importance of the various non-Gaussian contributions to the induced GW
spectrum, we now consider various primordial (Gaussian) curvature power spectra and present
numerical results for all terms.
In order to clearly illustrate the effects of non-Gaussianity, we typically fix the amplitude
$\mathcal{A}_\mathcal{R}$ and vary $F_\mathrm{NL}$.
Note, however, that when considering the production of PBHs, non-Gaussianity has a significant
impact on their abundance~\cite{Byrnes:2012yx,Young:2013oia,Franciolini:2018vbk}.
In addition, we display results for a range of parameters to study the contributions at each order
in $\mathcal{A}_\mathcal{R} F_\mathrm{NL}^2$ to the full GW spectrum.
In reality, for some of the most extreme values we consider, the perturbativity of the underlying
theory may break down (when $\mathcal{A}_\mathcal{R} F_\mathrm{NL}^2 \gtrsim 1$) or higher-order
terms in the expansion of the curvature perturbation (beyond the quadratic ansatz in \cref{eq:RNG})
may be important.
We retain these cases for illustrative purposes and to compare to existing literature, postponing
the consideration of higher-order corrections to future work.

\subsection{Monochromatic spectrum}

As a useful benchmark case, we first consider the spectrum of gravitational waves induced by a
monochromatic spectrum of density fluctuations,
\begin{align}\label{eqn:P-g-monochromatic}
    \Delta_g^2(k)
    &= \mathcal{A}_\mathcal{R}
        \delta( \ln \tilde{k} ),
\end{align}
defining $\tilde{k} = k / k_\star$.
The Gaussian result is~\cite{Kohri:2018awv}
\begin{align}
\begin{split}
   \Omega_\mathrm{GW}(k)_\mathrm{Gaussian}
    &= \frac{3 \mathcal{A}_\mathcal{R}^2}{1024}
        \tilde{k}^2
        \Theta(2 - \tilde{k})
        \left( \tilde{k}^2 - 4\right)^2
        \left( 3 \tilde{k}^2 - 2 \right)^2
    \\ &\hphantom{ {}={} }
        \times
        \left(
            \pi^2 \left( 3 \tilde{k}^2 - 2 \right)^2
            \Theta\left( 2 \sqrt{3} - 3 \tilde{k} \right)
            + \left[
                4
                + \left( 3 \tilde{k}^2 - 2 \right)
                \ln \abs{
                    \frac{4}{3 \tilde{k}^2} - 1
                }
            \right]^2
        \right).
\end{split}
\end{align}
Though the non-Gaussian terms must still be computed numerically, integrating over the Dirac delta
functions substantially reduces the dimensionality of the required integrals.
Note that for the nonplanar term, solving for the zeros of the Dirac delta functions requires
solving a quartic polynomial for one of the integration variables $s_i$ or $t_i$
(defined in \cref{app:recasting-the-integrals}).
In lieu of this we numerically integrate over one of the four Dirac delta functions, approximated as
a narrow lognormal function (\cref{eqn:P-g-lognormal} below).
We set $\sigma = 1/100$, which is more than sufficiently narrow to serve as a good approximation.

We begin by considering each non-Gaussian contribution individually.
\begin{figure}[t]
    \centering
    \includegraphics[width=\textwidth]
    {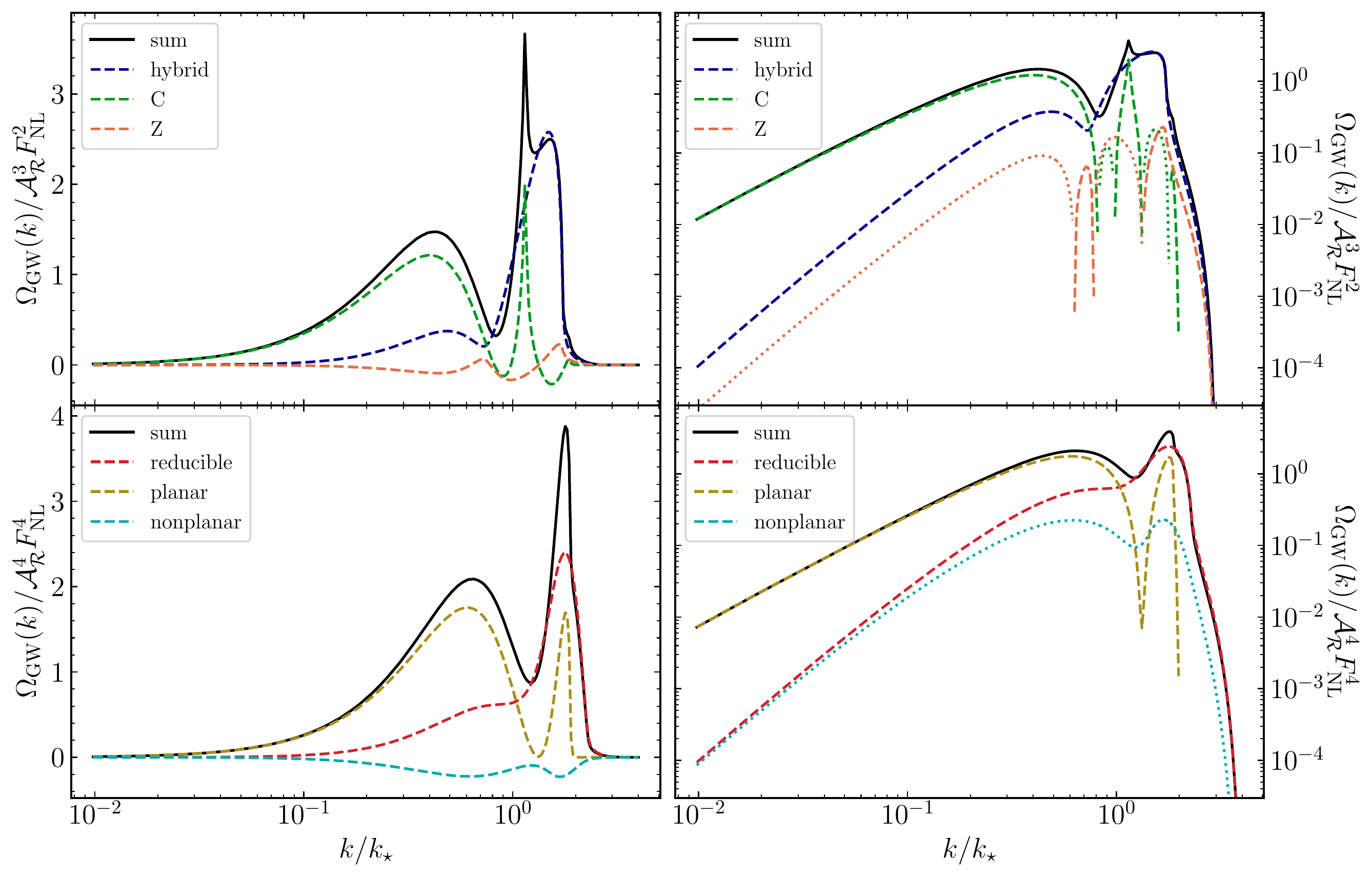}
    \caption{
        Unscaled (i.e., $\mathcal{A}_\mathcal{R} = 1$ and $F_\mathrm{NL} = 1$) non-Gaussian
        contributions for a monochromatic source, \cref{eqn:P-g-monochromatic}.
        The dashed lines indicate each contribution at $\mathcal{O}(F_\mathrm{NL}^2)$ (top panels)
        and at $\mathcal{O}(F_\mathrm{NL}^4)$ (bottom panels) with colors denoted in the legend.
        Solid black lines depict the sum of the individual contributions appearing in each panel.
        The left and right panels display the results on a linear and a log vertical scale,
        respectively.
        Because some individual contributions are negative at some $k$, dashed and dotted lines
        indicate where the sign is positive and negative, respectively.
    }
    \label{fig:monochromatic-ng-components}
\end{figure}
\Cref{fig:monochromatic-ng-components} displays the non-Gaussian contributions to the induced GW
spectrum, dividing the $\mathcal{O}(F_\mathrm{NL}^2)$ and $\mathcal{O}(F_\mathrm{NL}^4)$ terms by
$\mathcal{A}_\mathcal{R}^3 F_\mathrm{NL}^2$ and $\mathcal{A}_\mathcal{R}^4 F_\mathrm{NL}^4$,
respectively.
The connected terms clearly contribute as significantly as the disconnected ones, and they peak at
differing wavenumbers.
In particular, the \fourvertex{} and planar terms are substantially larger in the infrared and also
contribute comparably to the peaks at $k \gtrsim k_\star$.
In contrast, the \sunset{} and nonplanar terms are smaller in magnitude.
The \fourvertex{}, \sunset{}, and nonplanar terms are (for at least some $k$) negative; however, as
apparent in the right panels of \cref{fig:monochromatic-ng-components}, the summed contributions at
each order in $F_\mathrm{NL}$ are positive definite.

By comparing the peak heights of the $\mathcal{O}(F_\mathrm{NL}^2)$ and $\mathcal{O}(F_\mathrm{NL}^4)$
terms in the left panels of \cref{fig:monochromatic-ng-components}, we can estimate at what value of
$\mathcal{A}_\mathcal{R} F_\mathrm{NL}^2$ the two contributions are comparable.
For instance, aside from the spike in the \fourvertex{} term, the ratio of the peaks of the
$\mathcal{O}(F_\mathrm{NL}^2)$ and $\mathcal{O}(F_\mathrm{NL}^4)$ contributions is roughly
$1/2$.
In the common range $\mathcal{A}_\mathcal{R} \sim 10^{-2} - 10^{-3}$ considered for significant PBH
production, for $F_\mathrm{NL} \sim 5 - 20$ the $\mathcal{O}(F_\mathrm{NL}^4)$ terms contribute
significantly.
In the infrared (IR) limit, the ratios of the Gaussian and $\mathcal{O}(F_\mathrm{NL}^2)$
contributions is roughly $0.43$, while that for the $\mathcal{O}(F_\mathrm{NL}^2)$ and
$\mathcal{O}(F_\mathrm{NL}^4)$ ones is about $1.7$.

\begin{figure}[t]
    \centering
    \includegraphics[width=\textwidth]
    {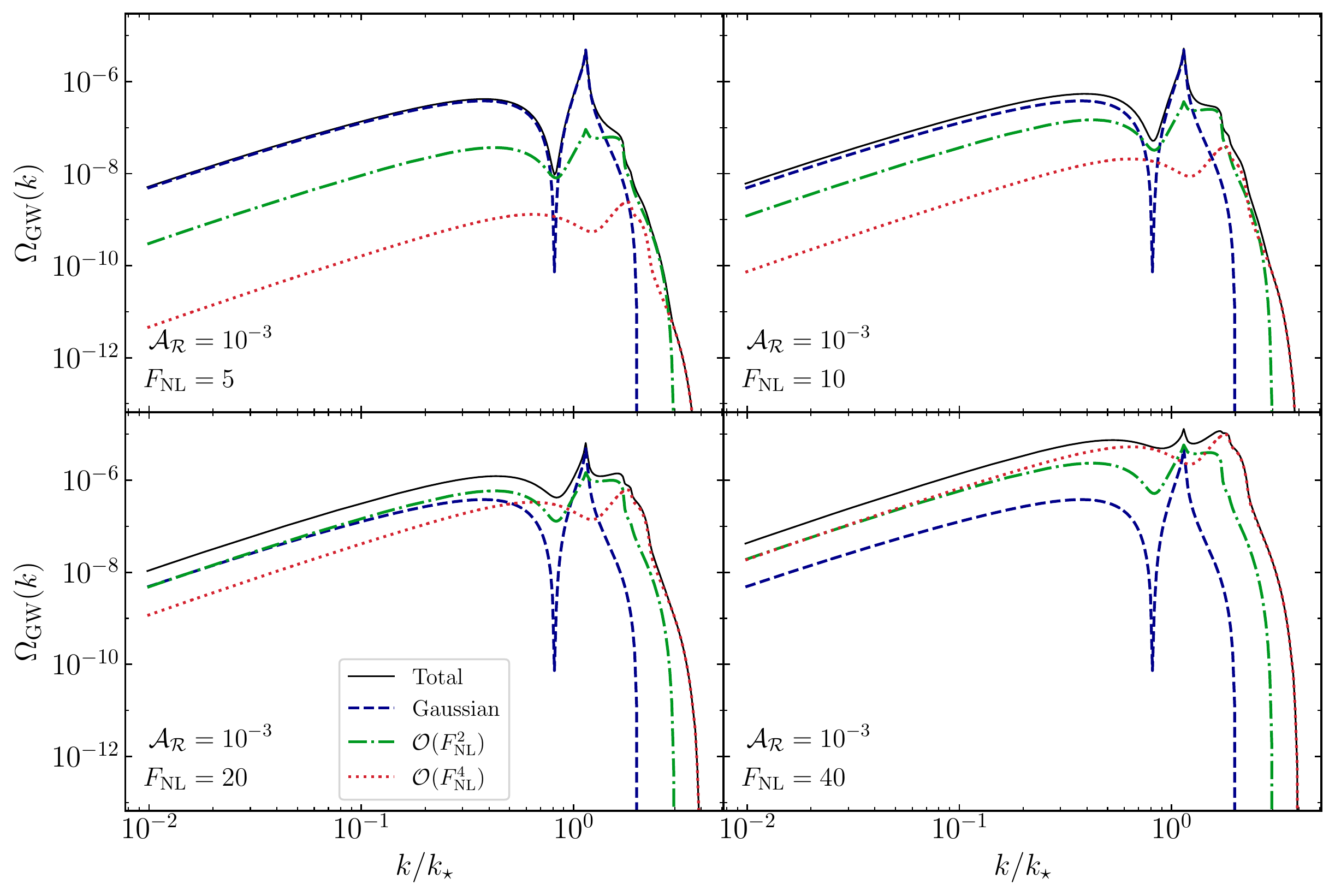}
    \caption{
        Induced GW spectrum after emission for a monochromatic source with $\mathcal{A}_\mathcal{R} =
        10^{-3}$ and various $F_\mathrm{NL}$.
        The Gaussian, $\mathcal{O}(F_\mathrm{NL}^2)$, and $\mathcal{O}(F_\mathrm{NL}^4)$ contributions
        appear in dashed blue, dot-dashed green, and dotted red respectively.
        The total spectrum is depicted in thin black.
    }
    \label{fig:monochromatic-gauss-vs-fnl2-vs-fnl4}
\end{figure}

We investigate the relative contributions of the $\mathcal{O}(F_\mathrm{NL}^2)$ and
$\mathcal{O}(F_\mathrm{NL}^4)$ terms in more detail in
\cref{fig:monochromatic-gauss-vs-fnl2-vs-fnl4}.
We fix $\mathcal{A}_\mathcal{R} = 10^{-3}$ and vary $F_\mathrm{NL}$ geometrically.
For $F_\mathrm{NL} = 5$ the Gaussian term dominates, but the non-Gaussian contributions produce
``knees'' near $k \sim 2 k_\star$ and $3 k_\star$ where the Gaussian contribution vanishes.
At $F_\mathrm{NL} = 10$ and $20$, the structure of the peak(s) is broadened by the
$\mathcal{O}(F_\mathrm{NL}^2)$ terms.
Finally, at $F_\mathrm{NL} = 40$ the $\mathcal{O}(F_\mathrm{NL}^2)$ and
$\mathcal{O}(F_\mathrm{NL}^4)$ terms contribute comparably and dominate over the Gaussian one,
resulting in a more complex peak.
Note that much of this structure is smoother in more realistic scenarios with broader scalar power
spectrum, as we investigate below.

As pointed out by Ref.~\cite{Yuan:2019wwo}, the infrared scaling of the induced GW spectrum includes
a logarithmic running on top of the typical pure power law behavior.
Though we verify that the spectral index of the disconnected non-Gaussian contributions may be
approximated as $3 + a \ln (b k)$ for some order-unity $a$ and $b$ (as found in
Ref.~\cite{Yuan:2020iwf}), of the connected terms this form only holds for the subdominant \sunset{}
and nonplanar ones.
The dominant non-Gaussian terms, the \fourvertex{} and planar, scale with $k^{2 + a \ln (b k)}$ like
the Gaussian term~\cite{Yuan:2019wwo}.\footnote{
    The infrared behavior of each contribution can be verified analytically by explicitly integrating
    over the Dirac delta functions and expanding in the limit $k / k_\star \ll 1$, a procedure we
    summarize in \cref{sec:monochromatic-analytic}.
}
As such, the spectral index of hypothetical GW signals could likely not be used to distinguish one
dominated by Gaussian vs. non-Gaussian contributions.
Furthermore, studies that neglect the connected terms significantly underestimate the non-Gaussian
contributions to the GW spectrum in the IR.

\begin{figure}[t]
    \centering
    \includegraphics[width=\textwidth]
    {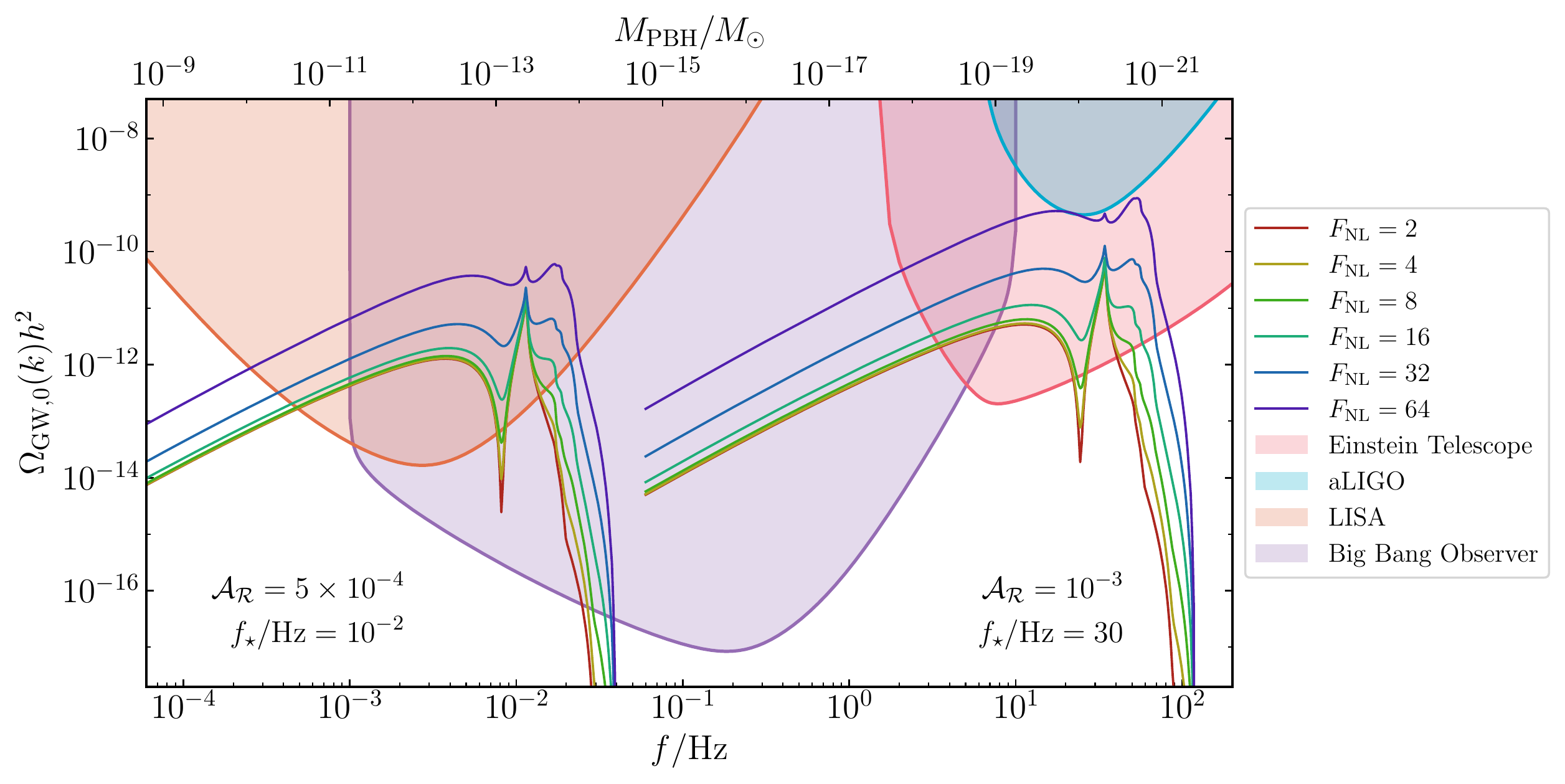}
    \caption{
        Present-day induced GW spectrum for a monochromatic source with various $F_\mathrm{NL}$,
        plotted against the sensitivity curves of various gravitational wave
        experiments~\cite{Schmitz:2020syl,schmitz_kai_2020_3689582}.
    }
    \label{fig:monochromatic-lisa-ligo-money}
\end{figure}

To make contact with potential observations, \cref{fig:monochromatic-lisa-ligo-money} depicts the
full, present-day GW signal alongside the sensitivity curves of various
experiments~\cite{Schmitz:2020syl}.\footnote{
    For simplicity, we compare to the power-law--integrated sensitivity curves provided by
    Ref.~\cite{Schmitz:2020syl}.
}
To get a sense of the sizes of PBHs that could possibly be produced in such scenarios, the top axis
of \cref{fig:monochromatic-lisa-ligo-money} depicts the mass $M_\mathrm{PBH}$ of PBHs produced by
the collapse of overdensities at horizon reentry on scales
$k_\star = 2 \pi f_\star$~\cite{Inomata:2016rbd},
\begin{align}
    \frac{M_\mathrm{PBH}}{M_\odot}
    &= \frac{\gamma}{0.2}
        \left( \frac{g_\star}{10.75} \right)^{-1/6}
        \left( \frac{f}{2.9 \times 10^{-9} \, \mathrm{Hz}} \right)^{-2},
\end{align}
taking $\gamma = 0.2$ and $g_\star = 106.75$ for frequencies in and above the LISA band.
We consider cases where $k_\star$ corresponds to a frequency of $10^{-2} \, \mathrm{Hz}$ in the LISA
band and $30 \, \mathrm{Hz}$ in the LIGO band.\footnote{
    Note that, for LIGO-band signals, if the primordial curvature spectrum were associated with PBH
    production, these PBHs (being lighter than $\sim 10^{-18} M_\odot$) would have evaporated via
    Hawking radiation by today~\cite{Hawking:1974rv, Hawking:1974sw}.
    In this work, however, we are agnostic as to the role the enhanced curvature spectrum plays in
    the production of PBHs.
}
One can observe how the signal shape changes with $F_\mathrm{NL}$, with a distinctive three-peak
structure arising at large $F_\mathrm{NL}$.

\subsection{Lognormal spectrum}\label{sec:lognormal}

\begin{figure}[ht]
    \centering
    \includegraphics[width=\textwidth]
    {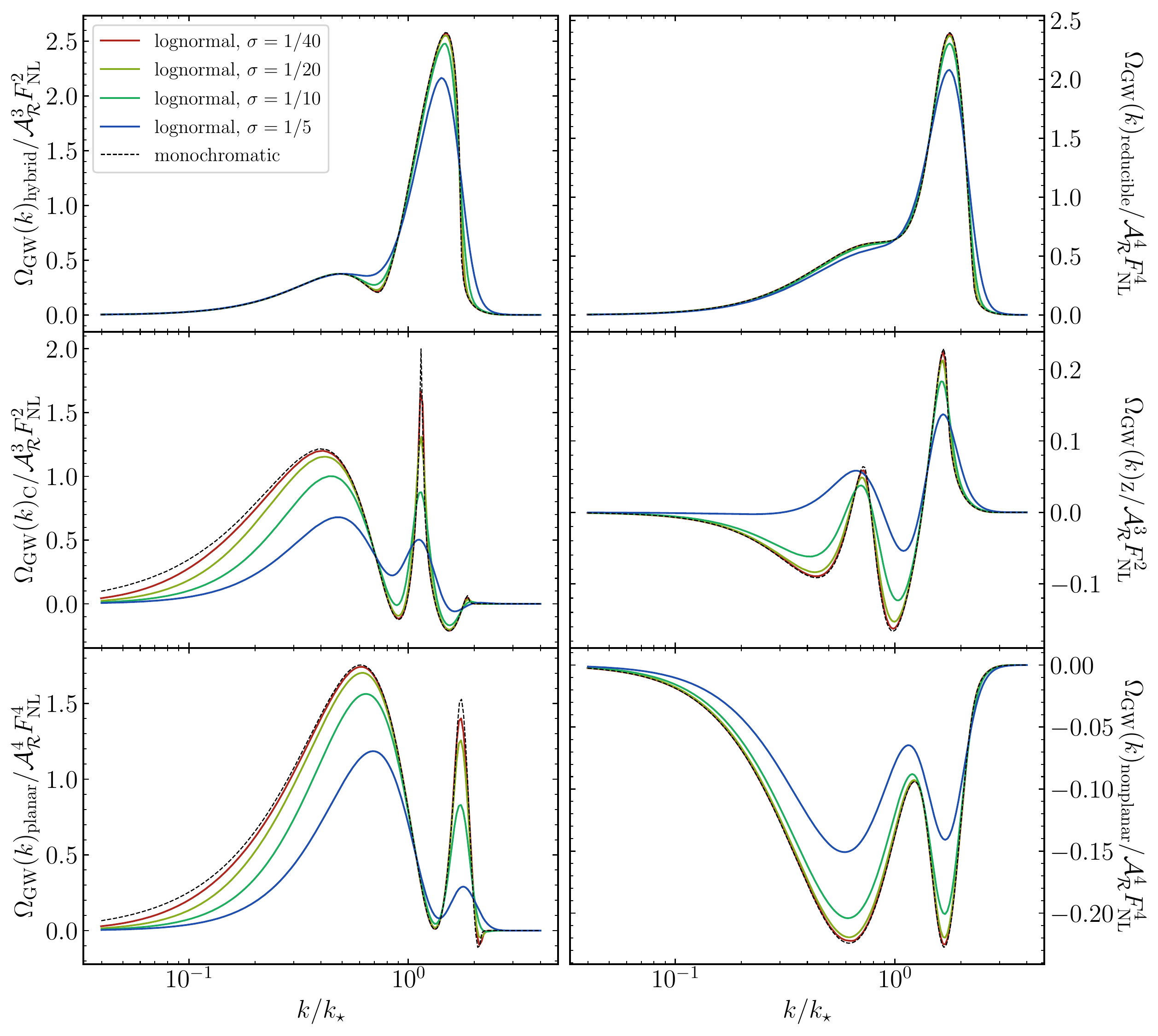}
    \caption{
        Unscaled (i.e., $\mathcal{A}_\mathcal{R} = 1$ and $F_\mathrm{NL} = 1$) non-Gaussian
        contributions for a lognormal source, \cref{eqn:P-g-lognormal}.
        Solid lines depict results for $\sigma$ spanning $1/40$ through $1/5$ as indicated
        in the legend.
        The monochromatic result is overlaid in thin, dashed black.
    }
    \label{fig:lognormal-components-vs-mono}
\end{figure}

We now explore how the features of the GW spectrum induced by an idealized, monochromatic source are
modulated when generalizing to a broader, more realistic spectrum.
Consider a spectrum with a Gaussian bump in $\ln(k)$, as in, e.g., Ref.~\cite{Unal:2018yaa},
\begin{align}\label{eqn:P-g-lognormal}
    \Delta_g^2(k)
    &= \frac{\mathcal{A}_\mathcal{R}}{\sqrt{2 \pi \sigma^2}}
        \exp\left( - \frac{ \ln^2 (k / k_\star)}{2 \sigma^2} \right),
\end{align}
normalized so that $\int \ud \ln k \, \Delta_g^2(k) = \mathcal{A}_\mathcal{R}$.
In \cref{fig:lognormal-components-vs-mono} we first study the effect of increasing $\sigma$ on each
of the individual non-Gaussian contributions in turn.
As one might expect, the various features in each contribution become less pronounced, shrinking in
amplitude and broadening in shape.
In addition, most contributions exhibit a $k^{3 + a \ln (b k)}$ scaling in the infrared.
Only for the Gaussian, C, and planar terms does an intermediate regime of $k^{2 + a \ln (b k)}$
behavior become partially evident for $\sigma = 1/40$, but each transitions to $k^{3 + a \ln (b k)}$
for $k / k_\star \lesssim 10^{-2}$.

\begin{figure}[ht]
    \centering
    \includegraphics[width=\textwidth]
    {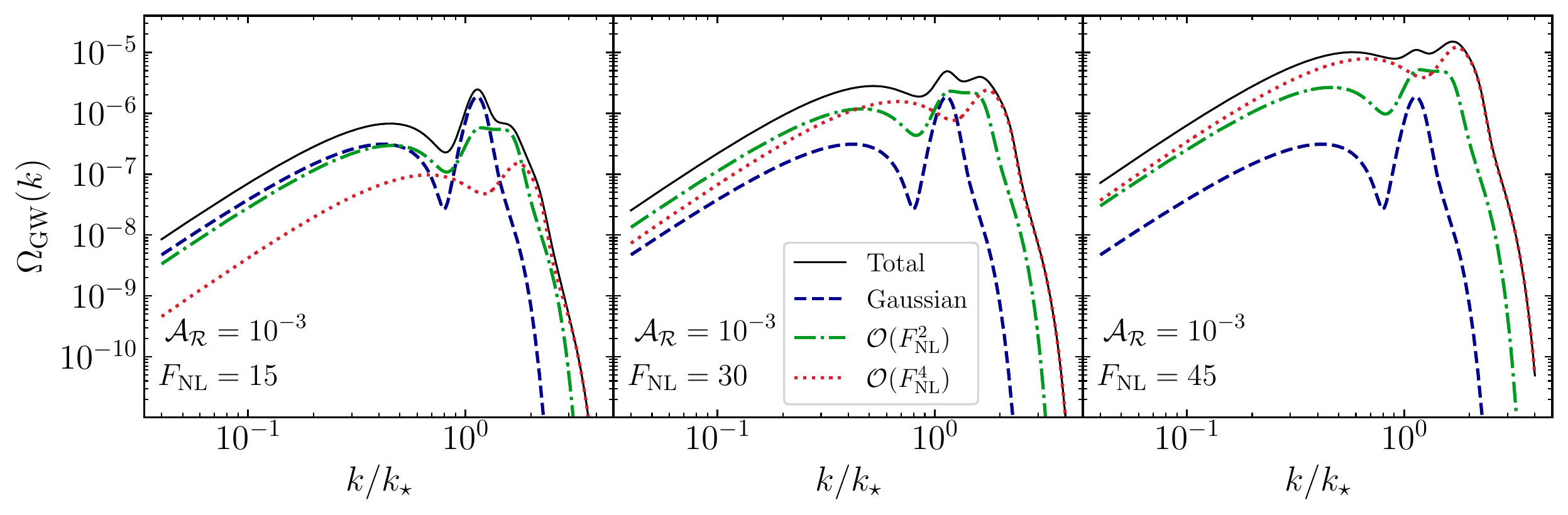}
    \caption{
        Induced GW spectrum after emission for a lognormal source with $\sigma = 1/10$,
        $\mathcal{A}_\mathcal{R} = 10^{-3}$ and various $F_\mathrm{NL}$.
        The Gaussian, $\mathcal{O}(F_\mathrm{NL}^2)$, and $\mathcal{O}(F_\mathrm{NL}^4)$ contributions
        appear in dashed blue, dot-dashed green, and dotted red respectively.
        The total spectrum is depicted in thin black.
    }
    \label{fig:lognormal-tenth-gauss-vs-fnl2-vs-fnl4}
\end{figure}

We investigate the relative contributions of the $\mathcal{O}(F_\mathrm{NL}^2)$ and
$\mathcal{O}(F_\mathrm{NL}^4)$ terms in more detail in
\cref{fig:lognormal-tenth-gauss-vs-fnl2-vs-fnl4}, taking $\sigma = 1/10$.
We again fix $\mathcal{A}_\mathcal{R} = 10^{-3}$ and vary $F_\mathrm{NL}$.
The peak structure is smoothed compared to that for the monochromatic spectrum,
\cref{fig:monochromatic-gauss-vs-fnl2-vs-fnl4}.
However, substantial non-Gaussianity does lead to a broad, nearly flat peak that distinguishes it
from the narrower feature evident in the spectrum for a purely Gaussian curvature perturbation.

\subsection{Gaussian-shaped spectrum}

\begin{figure}[t]
    \centering
    \includegraphics[width=\textwidth]{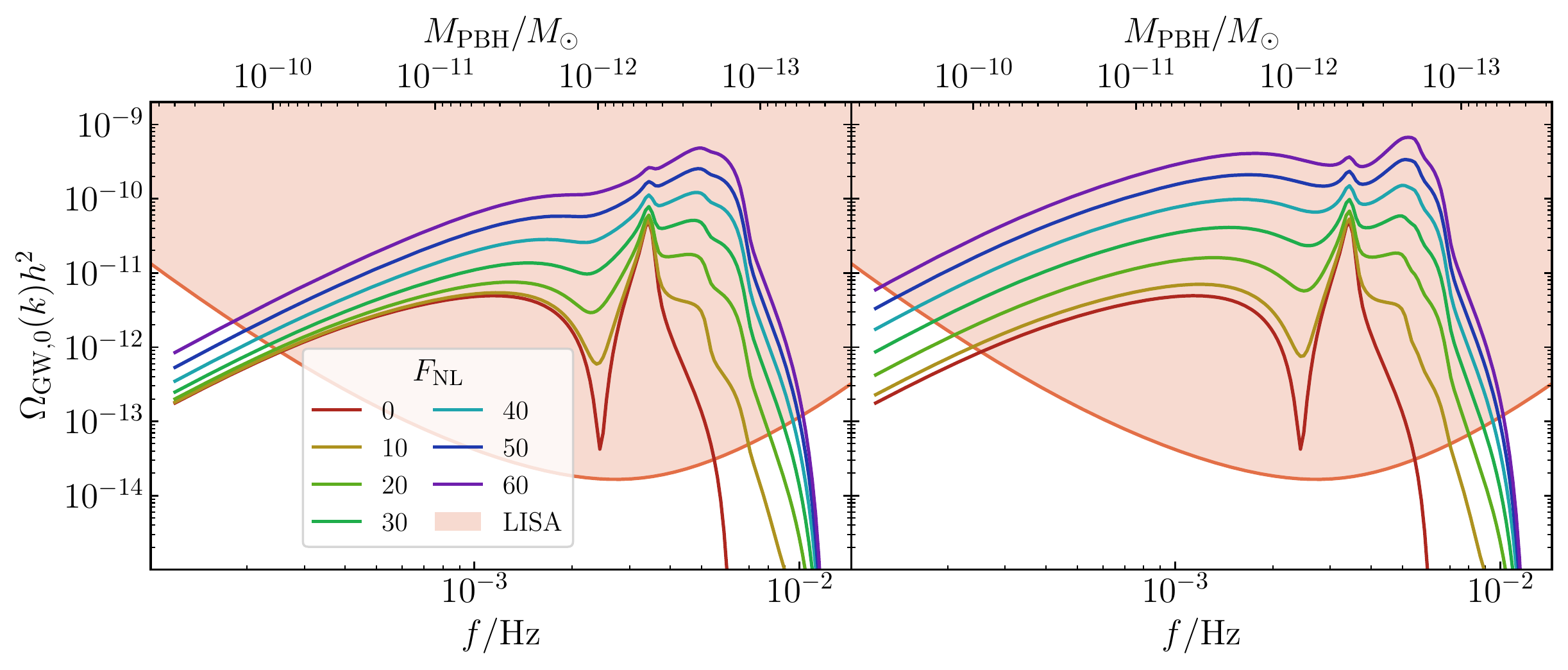}
    \caption{
        Gravitational wave spectra generated by a Gaussian-bump spectrum,
        \cref{eqn:P-g-gaussian-bump}, with $\sigma = k_\star/30$,
        $\mathcal{A}_\mathcal{R} = 10^{-3}$, and
        $k_\star / 2 \pi = 3 \times 10^{-3} \, \mathrm{Hz}$.
        The left and right panels exclude and include the connected contributions, respectively.
    }
    \label{fig:gaussian-peak}
\end{figure}

We next consider the Gaussian-bump spectrum used in Ref.~\cite{Cai:2018dig},
\begin{align}\label{eqn:P-g-gaussian-bump}
    \Delta_g^2(k)
    &= \left( \frac{k}{k_\star} \right)^3
        \frac{\mathcal{A}_\mathcal{R}}{\sqrt{2 \pi (\sigma / k_\star)^2}}
        \exp\left( - \frac{ (k - k_\star)^2}{2 \sigma^2} \right),
\end{align}
again normalized so that $\int \ud \ln k \, \Delta_g^2(k) = \mathcal{A}_\mathcal{R}$.
In \cref{fig:gaussian-peak} we show the total gravitational wave power spectrum for various
values of the non-Gaussianity parameter, $F_\mathrm{NL}$.
We compare results including and excluding the connected terms, choosing
$\sigma = k_\star/30$, $\mathcal{A}_\mathcal{R} = 10^{-3}$, and
$f_\star = k_\star / 2 \pi = 3 \times 10^{-3} \, \mathrm{Hz}$
to match the choices of Ref.~\cite{Cai:2018dig}.
Even when neglecting the connected terms, we do not reproduce the particular peak
structure observed in Ref.~\cite{Cai:2018dig}, and when including all contributions we observe
a more pronounced second peak around $2 f_\star$.
Though the peak amplitude near $f_\star$ is largely unchanged, neglecting the connected term
significantly underestimates the power at lower $k$ (as discussed in the monochromatic case above).

\subsection{Power law spectrum with an exponential cutoff}

Another common spectral shape is a power law that is exponentially cut off near some $k_\star$,
\begin{align}\label{eqn:plexp-spectrum}
    \Delta_g^2(k)
    &= \mathcal{A}_\mathcal{R} (k / k_\star)^{\alpha} e^{- \alpha (k / k_\star - 1)}.
\end{align}
This parameterization peaks at $k_\star$ with amplitude $\mathcal{A}_\mathcal{R}$ for any $\alpha$.
For example, Ref.~\cite{Inomata:2021uqj} found that, in contrast to the standard ultra slow roll
scenario, an inflationary potential with a small step could generate a curvature spectrum with
$\alpha = 4$ and a peak amplitude as large as $\mathcal{A}_\mathcal{R} \approx 10^{-2}$.
\Cref{eqn:plexp-spectrum} provides a good approximation to the result from
Ref.~\cite{Inomata:2021uqj}, aside from the oscillations in $k$.

\begin{figure}[t]
    \centering
    \includegraphics[width=\textwidth]
    {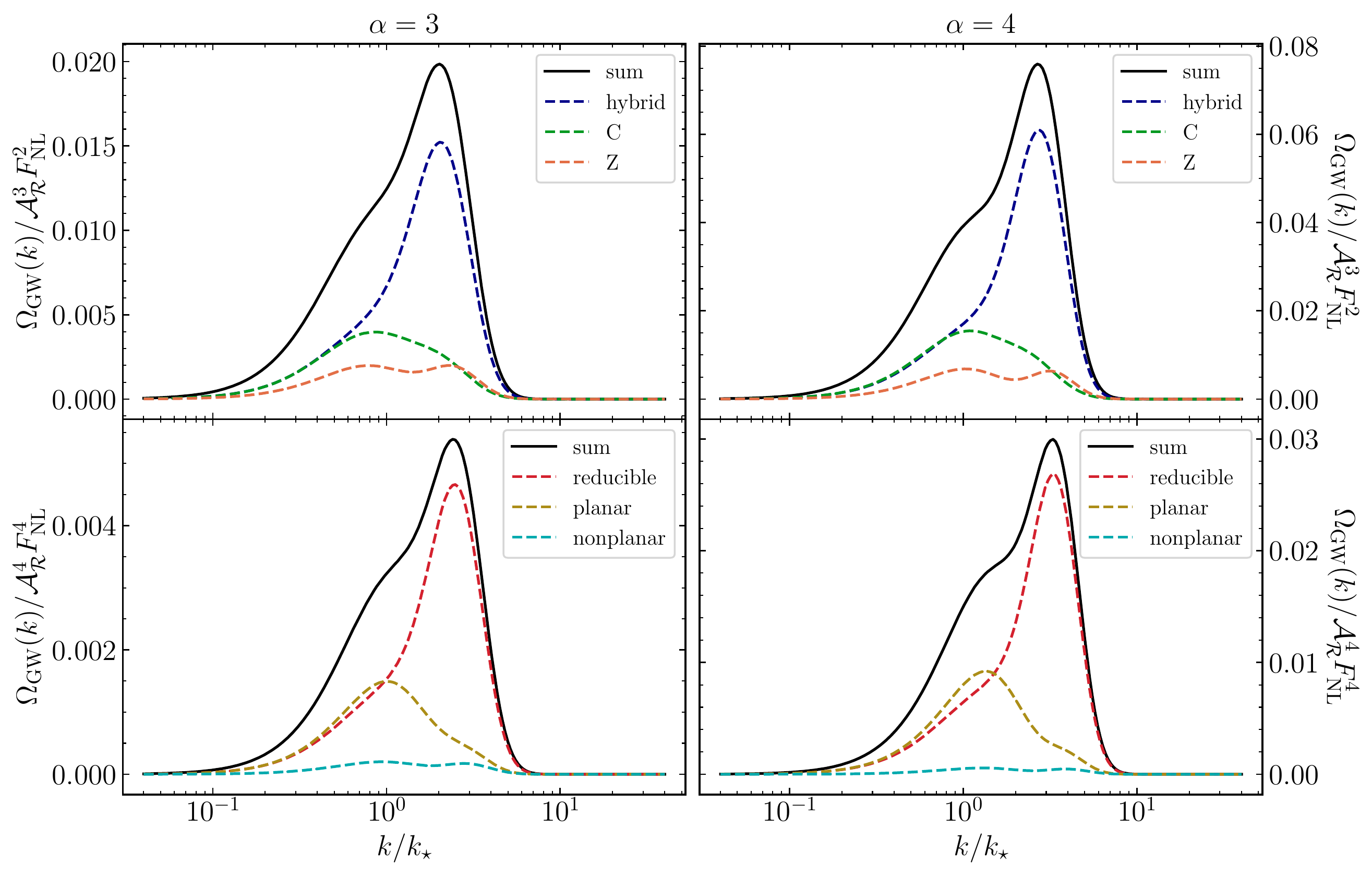}
    \caption{
        Unscaled (i.e., $\mathcal{A}_\mathcal{R} = 1$ and $F_\mathrm{NL} = 1$) non-Gaussian
        contributions for a power law source with an exponential cutoff, \cref{eqn:plexp-spectrum},
        for $\alpha = 3$ (left panels) and $4$ (right panels).
        The dashed lines indicate each contribution at $\mathcal{O}(F_\mathrm{NL}^2)$ (top panels)
        and at $\mathcal{O}(F_\mathrm{NL}^4)$ (bottom panels) with colors denoted in the legend.
        Solid black lines depict the sum of the individual contributions appearing in each panel.
    }
    \label{fig:plexp-3-and-4-ng-components}
\end{figure}

We display the individual results for each non-Gaussian term in
\cref{fig:plexp-3-and-4-ng-components}.
Like the monochromatic and lognormal sources, the connected and disconnected contributions peak at
differing $k / k_\star$, but their sum (at each order in $F_\mathrm{NL}$) does not exhibit as
prominent a multi-peak structure.
The results are also not highly sensitive to the value of $\alpha$.

\begin{figure}[ht]
    \centering
    \includegraphics[width=\textwidth]
    {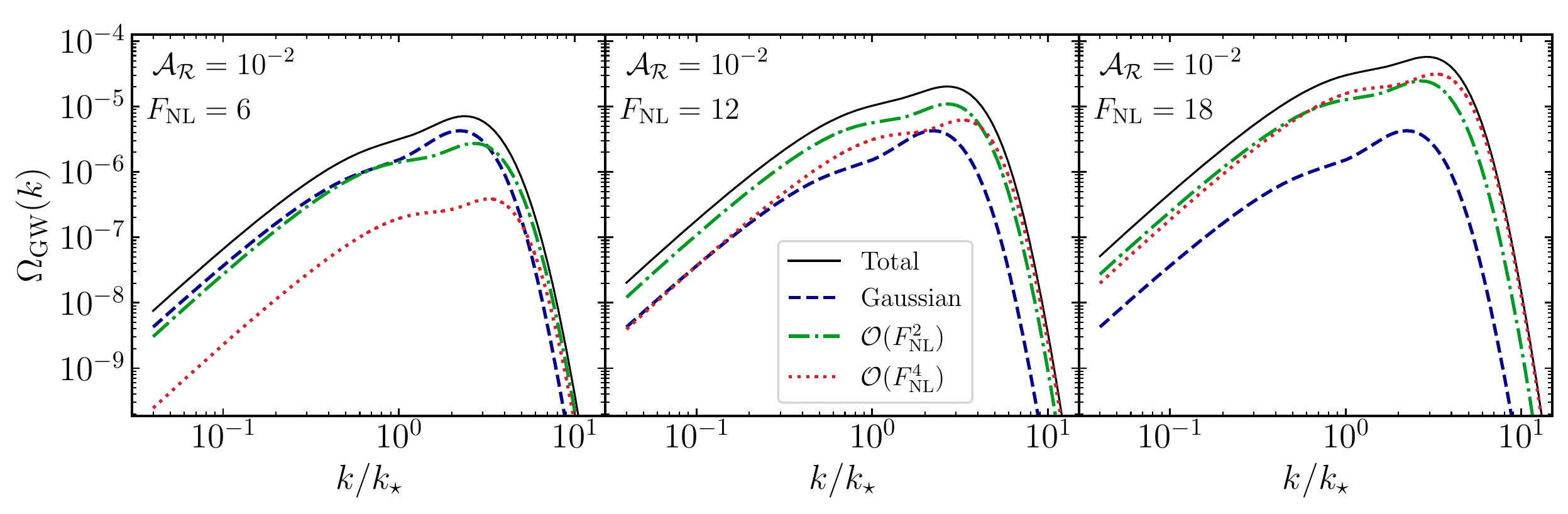}
    \caption{
        Induced GW spectrum after emission for a power law source with an exponential cutoff,
        \cref{eqn:plexp-spectrum}, with $\alpha = 4$,
        $\mathcal{A}_\mathcal{R} = 10^{-2}$ and various $F_\mathrm{NL}$.
        The Gaussian, $\mathcal{O}(F_\mathrm{NL}^2)$, and $\mathcal{O}(F_\mathrm{NL}^4)$ contributions
        appear in dashed blue, dot-dashed green, and dotted red respectively.
        The total spectrum is depicted in thin black.
    }
    \label{fig:plexp-4-gauss-vs-fnl2-vs-fnl4}
\end{figure}

We again depict the contributions of the $\mathcal{O}(F_\mathrm{NL}^2)$ and
$\mathcal{O}(F_\mathrm{NL}^4)$ terms for various $F_\mathrm{NL}$ in
\cref{fig:plexp-4-gauss-vs-fnl2-vs-fnl4} for $\alpha = 4$.
In contrast to sources that decay more quickly in the infrared, the shapes of the contributions to
different orders in $F_\mathrm{NL}$ are similar, each exhibiting a relatively broad peak and
an infrared scaling approaching $k^3$ with a moderate running.
From the left panel of \cref{fig:plexp-4-gauss-vs-fnl2-vs-fnl4} we may deduce that the
$\mathcal{O}(F_\mathrm{NL}^2)$ contributions are comparable to the Gaussian one when
$\mathcal{A}_\mathcal{R} F_\mathrm{NL}^2 \approx 0.3$, while the right panel indicates that the
$\mathcal{O}(F_\mathrm{NL}^2)$ and $\mathcal{O}(F_\mathrm{NL}^4)$ contributions match when
$\mathcal{A}_\mathcal{R} F_\mathrm{NL}^2 \approx 3$.
However, the signal is cut off at larger $k / k_\star$ depending on whether (and which of) the
non-Gaussian terms dominate.

\begin{figure}[ht]
    \centering
    \includegraphics[width=.8\textwidth]
    {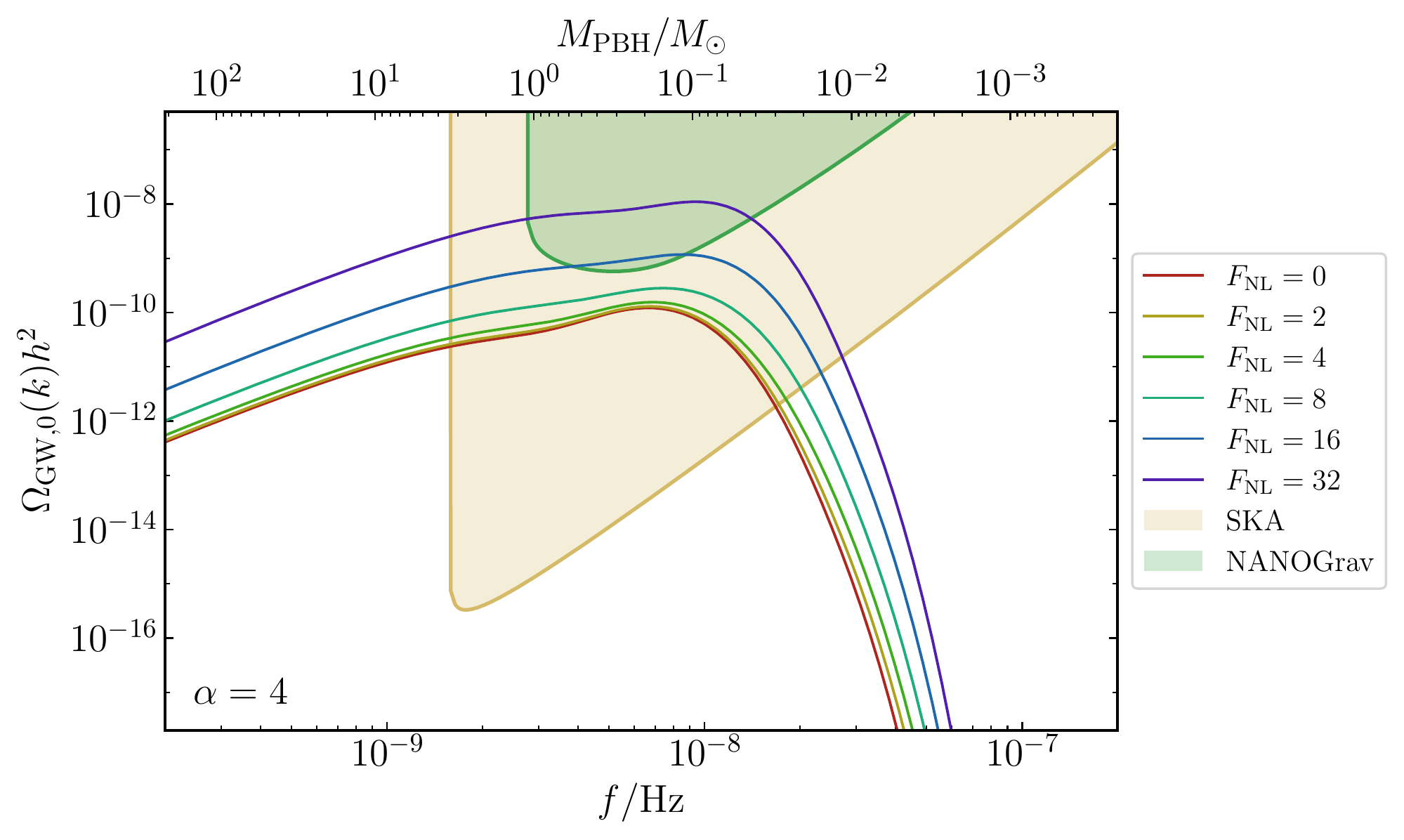}
    \caption{
        Present-day induced GW spectrum for a power law ($\alpha = 4$) source with an exponential
        cutoff, setting $\mathcal{A}_\mathcal{R} = 10^{-2}$.
        The shaded regions indicate the sensitivity curves for the pulsar timing experiments
        NANOGrav and the Square Kilometer Array (SKA)~\cite{Schmitz:2020syl,schmitz_kai_2020_3689582}.
    }
    \label{fig:plexp-pta-money}
\end{figure}

Finally, we present results for the benchmark scenario of Ref.~\cite{Inomata:2021uqj} (for the
Gaussian part, including the effects of additional local-Gaussianity for illustrative purposes) in
\cref{fig:plexp-pta-money}, with a peak frequency of $k_\star / 2 \pi = 3 \, \mathrm{nHz}$ and
amplitude $\mathcal{A}_\mathcal{R} = 10^{-2}$ (and $g_\star = 10.75$).

\subsection{Broken power law spectrum}

We next study a broken power law spectrum, as considered in Ref.~\cite{Atal:2021jyo}, with
parameterization
\begin{align}
    \Delta_g^2(k)
    &= \frac{\mathcal{A}_\mathcal{R}}{(k / k_\star)^{- \alpha} + (k / k_\star)^\beta},
\end{align}
scaling with $k^\alpha$ in the IR and $k^{-\beta}$ in the ultraviolet (UV).
All contributions to the GW power exhibit a spectral index of $3+a\ln (bk)$ in the infrared as in
the lognormal case in \Cref{sec:lognormal}.
In the ultraviolet the spectral index approaches $- 2 \beta$ for $0 < \beta < 4$ and
$-4 - \beta$ for $\beta > 4$.
These scalings agree with Ref.~\cite{Atal:2021jyo}, which provided estimates of the IR and UV
behavior for the hybrid and \sunset{} contributions.
Aside from decaying as a power law in the UV rather than exponentially, the qualitative features of
the induced GW spectrum for this case are similar to that for the exponentially cut off power law.

\section{Conclusions}\label{sec:conclusion}

In this work we have carefully computed the effect of local non-Gaussianity on the spectrum of
gravitational waves induced by scalar fluctuations.
At lowest order in fluctuations, the induced gravitational wave spectrum is sourced by the
trispectrum, or 4-point function of curvature fluctuations.
We have shown that, contrary to some previous studies, the connected part of the trispectrum makes
important contributions to the total spectrum that can neither be entirely neglected nor
approximated by a multiple of the disconnected contributions.
Our results demonstrate that studies of the induced GW spectrum from non-Gaussian curvature
perturbations must carefully consider (and compute) all such contributions.
For the power spectrum enhancements that are commonly considered in PBH scenarios
(with $\mathcal{A}_\mathcal{R} \sim 10^{-3}$), even modest values of
$F_\mathrm{NL} \sim \mathcal{O}(1)$ to $\mathcal{O}(10)$ significantly impact the induced GW
spectrum.

A number of possible extensions merit attention.
We have considered only the standard local-type non-Gaussianity, while more generally other
bispectrum shapes may be relevant in detailed constructions, such as Ref.~\cite{Inomata:2021uqj}.
Furthermore, connected contributions must also be considered for higher-order non-Gaussianity, such
as that considered in Ref.~\cite{Yuan:2020iwf}.
We leave these considerations for future work.

\acknowledgments

We gratefully thank Keisuke Inomata for comments on a draft of this manuscript and Caner Unal and
Guillem Domenech for useful discussions.
This work was supported in part by the US Department of Energy through grant DESC0015655.
Z.J.W.\ is supported in part by the United States Department of Energy Computational Science
Graduate Fellowship, provided under Grant No. DE-FG02-97ER25308.
This work made use of the Illinois Campus Cluster, a computing resource that is operated by the
Illinois Campus Cluster Program (ICCP) in conjunction with the National Center for Supercomputing
Applications (NCSA) and which is supported by funds from the University of Illinois at
Urbana-Champaign.
This work made use of the Python packages \textsf{vegas}~\cite{Lepage:2020tgj},
\textsf{NumPy}~\cite{harris2020array}, \textsf{SciPy}~\cite{scipy},
\textsf{matplotlib}~\cite{matplotlib}, and \textsf{emcee}~\cite{ForemanMackey:2012ig}.

\appendix

\section{Cosmological perturbation theory}\label{app:cpt}

In this appendix we elaborate on the dynamics of cosmological perturbations and the gravitational
wave spectrum induced at second order by scalar perturbations.
Recall that the background Einstein equations set
\begin{subequations}\label{eqn:background-einstein-eqns}
\begin{align}
    \mathcal{H}^2
    &= \frac{a^2}{3 \mpl^2} \overbar{\rho} \\
	\mathcal{H}' + \mathcal{H}^2
    &= \frac{a^2}{6 \mpl^2} \left( \overbar{\rho} - 3 \overbar{P} \right).
\end{align}
\end{subequations}
We define the perturbations to the stress-energy tensor as
\begin{subequations}
\label{eqn:stress-tensor-decomp-mixed}
\begin{align}
    \delta T^{0}_{\hphantom{0}0}
    &= - \delta \rho \\
    \delta T^{0}_{\hphantom{0}i}
    &= \left( \overbar{\rho} + \overbar{P} \right)
        \partial_i \delta u \\
    \label{eqn:stress-tensor-decomp-mixed-Ti0}
    \delta T^{i}_{\hphantom{i}j}
    &= \delta_{ij} \delta P,
\end{align}
\end{subequations}
neglecting vector and tensor perturbations and scalar anisotropic stress.
We can solve the first-order Einstein equations for $\delta \rho$, $\delta P$, and $\delta u$ as
\begin{subequations}
\label{eqn:scalar-flucs-ito-metric}
\begin{align}
    \delta \rho
    &= - \frac{2 \mpl^2}{a^2} \left(
            3 \mathcal{H} \left( \Phi' + \mathcal{H} \Phi \right)
            - \partial_i \partial_i \Phi
        \right) \\
    \delta P
    &= \frac{6 \mpl^2}{a^2}
        \left(
            \Phi'' + 3 \mathcal{H} \Phi'
         \right)
         - 6 \overbar{P} \Phi \\
    \delta u
    &= - \frac{2 \mpl^2}{a^2 \left( \overbar{\rho} + \overbar{P} \right)}
        \left( \Phi' + \mathcal{H} \Phi \right).
\end{align}
\end{subequations}

The terms in the space-space components of the Einstein tensor that are second order in the
Newtonian potential are
\begin{align}
    a^2 \delta^{(2)} G_i^{\hphantom{i}j}
    &= 4 \Phi \partial_i \partial_j \Phi
        + 2 \partial_i \Phi \partial_j \Phi
        + \delta_i^{\hphantom{i}j} \left(
            20 \left( \mathcal{H}^2 + \mathcal{H}' \right) \Phi^2
            + 8 \mathcal{H} \Phi \Phi'
            - 4 \Phi \partial_k \partial_k \Phi
            - {\Phi'}^2
            - 3 \partial_k \Phi \partial_k \Phi
        \right).
\end{align}
GWs are also sourced by second-order perturbations to the stress-energy tensor, which, using
\cref{eqn:scalar-flucs-ito-metric}, may be expressed in terms of metric perturbations as
\begin{align}
    \delta^{(2)} T_i^{\hphantom{i}j}
    &= \frac{4 \mpl^4}{a^4 \left( \overbar{\rho} + \overbar{P} \right)}
        \partial_i \left( \Phi' + \mathcal{H} \Phi \right)
        \partial_j \left( \Phi' + \mathcal{H} \Phi \right).
\end{align}
Dropping terms proportional to $\delta_{ij}$, setting $\overbar{P} = w \overbar{\rho}$, and
substituting \cref{eqn:background-einstein-eqns},
\begin{align}
    \mathcal{S}_{ij}
    &= 4 \Phi \partial_i \partial_j \Phi
        + \frac{2 (1 + 3 w)}{3 (1 + w)} \partial_i \Phi \partial_j \Phi
        - \frac{4}{3 (1 + w) \mathcal{H}^2}
        \left[
            \partial_i \Phi' \partial_j \Phi'
            + \mathcal{H} \partial_i \Phi \partial_j \Phi'
            + \mathcal{H} \partial_i \Phi' \partial_j \Phi
        \right].
    \label{eqn:Sij-second-order}
\end{align}

After using \cref{eqn:transfer-times-curvature} to express the Newtonian potential in terms of the
comoving curvature perturbation $\mathcal{R}(\mathbf{k})$ and the transfer function $\Phi(k \tau)$,
taking a Fourier transform, and projecting onto (negative) $\epsilon^\lambda_{ij}(\mathbf{k})$,
we find
\begin{align}
    \mathcal{S}_\lambda(\tau, \mathbf{k})
    &\equiv - \epsilon^\lambda_{lm}(\mathbf{k}) \mathcal{S}_{lm}(\mathbf{k}) \\
    &= \int \frac{\ud^3 q}{(2 \pi)^{3/2}}
        Q_\lambda(\mathbf{k}, \mathbf{q})
        f(\abs{\mathbf{k} - \mathbf{q}}, q, \tau)
        \mathcal{R}(\mathbf{q}) \mathcal{R}(\mathbf{k} - \mathbf{q}),
    \label{eqn:source-definition}
\end{align}
with
\begin{align}\label{eqn:def-f-general}
\begin{split}
    f(p, q, \tau)
    &\equiv
        \frac{3 (1 + w)}{(5 + 3 w)^2}
        \bigg[
            2 (5 + 3 w)
            \Phi(p \tau) \Phi(q \tau)
            + \frac{4}{\mathcal{H}^2} \Phi'(p \tau) \Phi'(q \tau)
    \\ &\hphantom{ {}={} \frac{3 (1 + w)}{(5 + 3 w)^2} \Bigg[ }
            + \frac{4}{\mathcal{H}} \left(
                \Phi(p \tau) \Phi'(q \tau) + \Phi'(p \tau) \Phi(q \tau)
            \right)
        \bigg].
\end{split}
\end{align}
Substituting $\mathcal{H} = \alpha / \tau$ (via \cref{eqn:scale-factor-fixed-eos}) yields
\cref{eqn:source-function-p-q-tau}.
The projection factors $Q_\lambda$ defined in \cref{eqn:projection-factor-def} obey the symmetries
\begin{subequations}\label{eqn:projection-factor-symmetries}
\begin{align}
    Q_{\lambda}(\mathbf{k}, \mathbf{q})
    &= Q_{\lambda}(\mathbf{k}, \mathbf{q} + \gamma\mathbf{k}) \\
    Q_{\lambda}(\mathbf{k}, \mathbf{q})
    &= Q_{\lambda}(-\mathbf{k}, \mathbf{q})
    = Q_{\lambda}(\mathbf{k}, -\mathbf{q})
    = Q_{\lambda}(-\mathbf{k}, -\mathbf{q}).
\end{align}
\end{subequations}

We next require the solution to the equation of motion of the transfer function, $\Phi(k \tau)$.
In the absence of isocurvature perturbations, the Newtonian potential evolves according to
\begin{align}\label{eqn:newtonian-potential-eom}
    \Phi''(\tau, \mathbf{k})
        + 3 (1 + w) \mathcal{H} \Phi'(\tau, \mathbf{k})
        + w k^2 \Phi(\tau, \mathbf{k})
    &= 0,
\end{align}
after setting $\delta P = w \delta \rho$.
Using \cref{eqn:scale-factor-fixed-eos}, \cref{eqn:newtonian-potential-eom} leads to
\begin{align}\label{eqn:transfer-eom-ito-alpha}
    0
    &= \frac{\ud^2 \Phi}{\ud y^2}
        + \frac{6 (1 + w)}{1 + 3 w} \frac{1}{y} \frac{\ud \Phi}{\ud y}
        + \Phi,
\end{align}
where $y \equiv \sqrt{w} k \tau$.
For $w \neq 0$, the solutions are given in terms of the spherical Bessel functions of the first and
second kind, $j_\nu$ and $y_\nu$:
\begin{align}
    \Phi(k \tau)
    &= \frac{
            C_1 j_{\gamma - 1}(y)
            + C_2 y_{\gamma - 1}(y)
        }{
            y^{\gamma - 1}
        },
\end{align}
where $\gamma = 3 (1 + w) / (1 + 3 w)$.
Imposing the superhorizon initial conditions $\Phi(x \to 0) = 1$ and $\partial_x \Phi(x \to 0) = 0$
sets $C_1 = 1$ and $C_2 = 0$.
In radiation domination, $\gamma = 2$ and
\begin{align}
    \Phi(y)
    &= \frac{\sin y - y \cos y}{y^3}
\end{align}
In matter domination, the solution to \cref{eqn:newtonian-potential-eom} for the transfer function
is instead merely $\Phi(k \tau) = 1.$

The Green function for the tensor perturbations (i.e., the solution to \cref{eqn:green-function-eom})
is constructed via
\begin{align}
    G_\mathbf{k}(\tau, \overbar{\tau})
    &= \frac{
            v_1(k \tau) v_2(k \overbar{\tau})
            - v_1(k \overbar{\tau}) v_2(k \tau)
        }{
            v_1'(k \overbar{\tau}) v_2(k \overbar{\tau})
            - v_1(k \overbar{\tau}) v_2'(k \overbar{\tau})
        },
\end{align}
where $v_1$ and $v_2$ are the homogeneous solutions to the equation of motion for
$v(k \tau) = a(\tau) h_\lambda(\tau, \mathbf{k})$,
\begin{align}
    \frac{\ud^2 v}{\ud x^2}
    + \left( 1 - \frac{\alpha (\alpha - 1)}{x^2} \right) v
    = 0.
\end{align}
The solutions are
\begin{align}\label{eqn:homogeneous-solutions-v}
    v_1(x)
    &= x j_{\alpha - 1}(x)
    \\
    v_2(x)
    &= x y_{\alpha - 1}(x),
\end{align}
leading to
\begin{align}
    G_\mathbf{k}(\tau, \overbar{\tau})
    &= k \tau \overbar{\tau} \left[
            j_{\alpha - 1}(k \overbar{\tau}) y_{\alpha - 1}(k \tau)
            - j_{\alpha - 1}(k \tau) y_{\alpha - 1}(k \overbar{\tau})
        \right].
\end{align}
In radiation domination where $w = 1/3$ and so $\alpha = 1$, the Green function is
\begin{align}
    G_\mathbf{k}(\tau, \overbar{\tau})
    = \frac{\sin k (\tau - \overbar{\tau})}{k}.
\end{align}

In, e.g., radiation domination, \cref{eqn:I-def} can be computed analytically in terms of cosine and
sine integrals via repeated integration by parts~\cite{Kohri:2018awv}.
Defining $\tilde{I}(v, u, x) \equiv k^2 I(v k, u k, x / k)$, the quantity required to compute the
observable gravitational wave spectrum is
\begin{align}\label{eqn:late-time-transfer-function-product}
\begin{split}
    \overbar{\tilde{I}(v_1, u_1, x \to \infty) \tilde{I}(v_2, u_2, x \to \infty)}
    &= \frac{1}{2 x^2}
        \tilde{I}_A(u_1, v_1)
        \tilde{I}_A(u_2, v_2)
        \Big[
            \tilde{I}_B(u_1, v_1)
            \tilde{I}_B(u_2, v_2)
    \\ &\hphantom{
            {}={}
            \frac{1}{2 x^2}
            \tilde{I}_A(u_1, v_1)
            \tilde{I}_A(u_2, v_2)
            \Big[
        }
            + \pi^2
            \tilde{I}_C(u_1, v_1)
            \tilde{I}_C(u_2, v_2)
        \Big],
\end{split}
\end{align}
where
\begin{subequations}
\begin{align}
    \tilde{I}_A(u, v)
    &= \frac{3 (u^2 + v^2 - 3)}{4 u^3 v^3} \\
    \label{eqn:I-B}
    \tilde{I}_B(u, v)
    &= - 4 u v
        + (u^2 + v^2 - 3) \ln\abs{\frac{3 - (u + v)^2}{3 - (u - v)^2}} \\
    \tilde{I}_C(u, v)
    &= \left( u^2 + v^2 - 3 \right)
        \Theta(v + u - \sqrt{3}).
\end{align}
\end{subequations}

\section{Diagrammatic rules for the gravitational wave spectrum}
\label{app:trispectrum}

In this appendix, we detail the computation of the contributions to the primordial trispectrum that
induce gravitational waves.
While a direct computation is straightforward, if tedious, we also present a diagrammatic
representation of the non-Gaussian contributions.
In order to make contact with existing literature, we present explicit Feynman-type rules with which
one can represent each of the integrals contributing to the induced GW spectrum.

A complication in this approach is that the transfer functions relate the (nonlinear) curvature
perturbation amplitudes to the Newtonian potential.
The momenta flowing through the transfer functions must be tracked, and as we demonstrate below this
leads to differences in diagrams that are otherwise topologically identical.
We denote transfer functions with dashed lines; 3-momenta flow through these in the direction
indicated by the arrow.
The rules are given in \cref{Tab:DiagramRules}.
They function like regular Feynman rules: one draws all allowed diagrams and integrates over all
loop momenta, momentum is conserved at each vertex, and the overall momentum of a diagram is zero.
\begin{table}[th]
\begin{center}
\begin{tabular*}{\textwidth}{c @{\extracolsep{\fill}} c @{\extracolsep{\fill}} c}
    (i)
    & \raisebox{-.5\height}{\includegraphics[width=.18\textwidth]{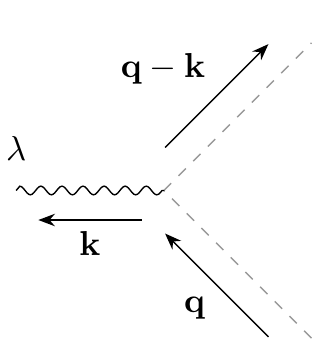}}
    & $\displaystyle 4 \int_{\tau_0}^{\tau} \ud \overbar{\tau} \,
        \frac{a(\overbar{\tau})}{a(\tau)}
        G_\mathbf{k}(\tau, \overbar{\tau})
        Q_\lambda(\mathbf{k}, \mathbf{q})
        f(\abs{ \mathbf{k} - \mathbf{q} }, q, \overbar{\tau})$
    \\
    (ii)
    & \raisebox{-.5\height}{\includegraphics[width=.18\textwidth]{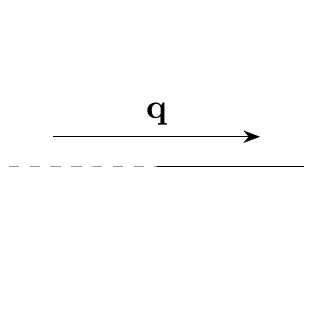}}
    & $1$
    \\
    (iii)
    & \raisebox{-.5\height}{\includegraphics[width=.18\textwidth]{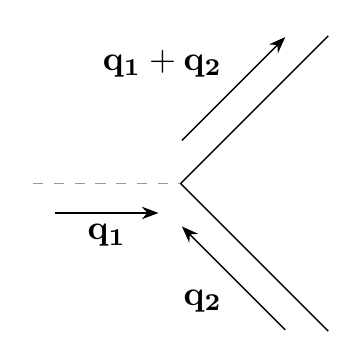}}
    & $F_\mathrm{NL}$
    \\
  (iv)
    & \raisebox{-.5\height}{\includegraphics[width=.18\textwidth]{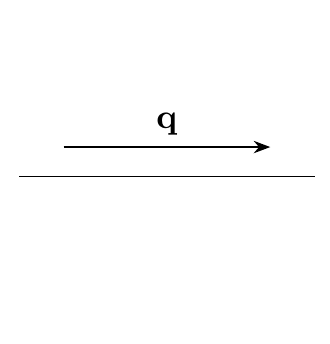}}
    & $\mathcal{P}_g(q)$
\end{tabular*}
\caption{
    Rules for the diagrammatic representation of Gaussian and local $F_\mathrm{NL}$-type
    non-Gaussian contributions to the induced GW spectrum.
    Wavy lines are gravitational waves, solid lines are (Gaussian) scalar power spectra, and dashed
    lines represent the transfer function of the Newtonian potential.
    Diagrams that include vertices in rule (iii) where the solid lines are connected into a loop
    vanish by virtue of the definition of $\mathcal{R}$ in \cref{eq:RNG}.
}\label{Tab:DiagramRules}
\end{center}
\end{table}
Note that, even when performing the computations algebraically, a diagrammatic representation that
accounts for the transfer functions facilitates determining the multiplicity of each term (while
keeping in mind the symmetries of the transfer function and \cref{eqn:projection-factor-symmetries}).

\begin{figure}[t]
   \centering
   \includegraphics[width=.48\textwidth]{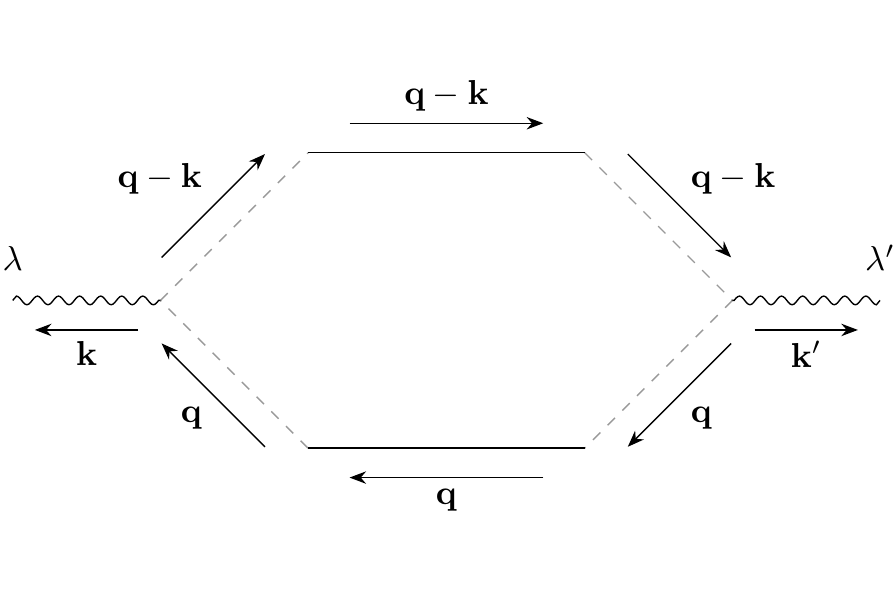}
   \caption{
        The Gaussian diagram, contributing at $\mathcal{O}(F_\mathrm{NL}^0)$ to the GW power spectrum.
    }
   \label{fig:FNL0Feynman}
\end{figure}

\begin{figure}[t]
   \centering
   \includegraphics[width=.48\textwidth]{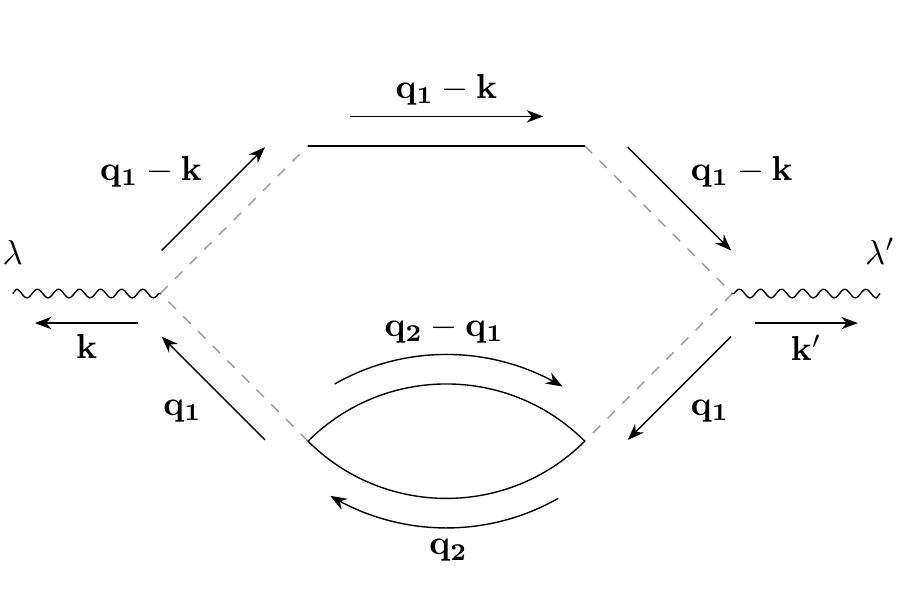}\hfill
   \includegraphics[width=.48\textwidth]{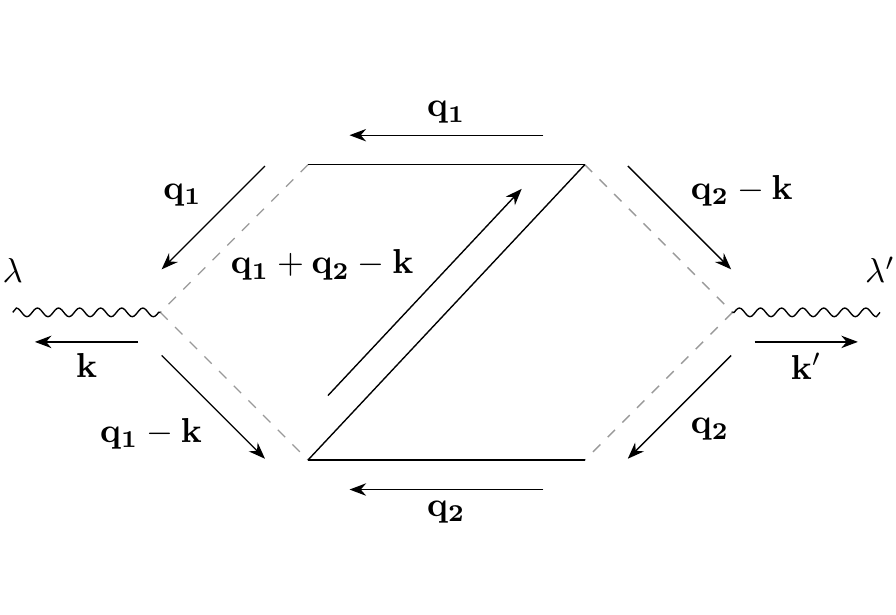}
   \includegraphics[width=.48\textwidth]{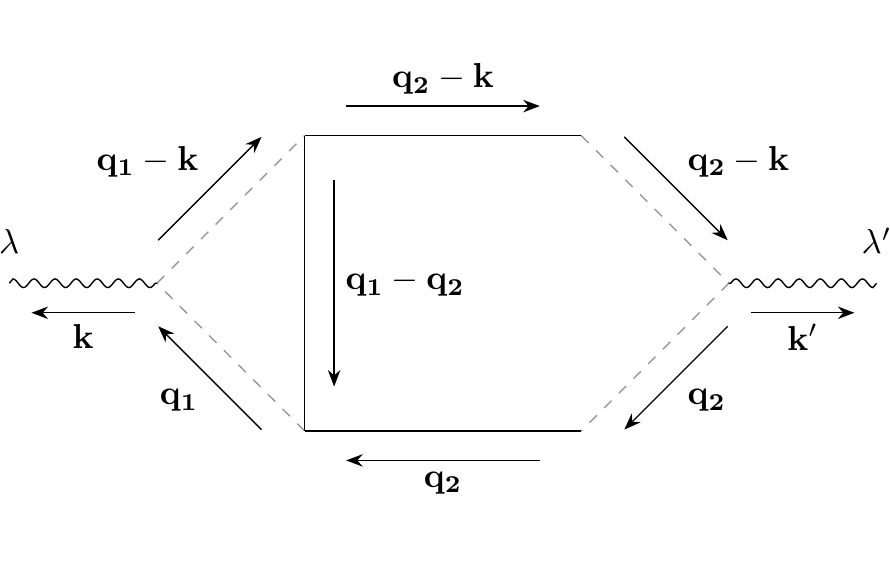}\hfill
   \includegraphics[width=.48\textwidth]{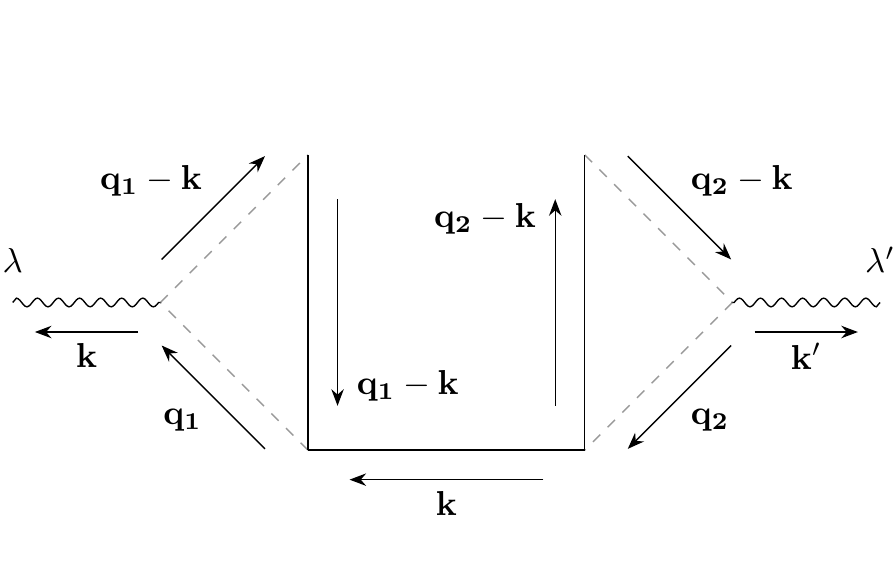}
   \caption{
       Hybrid (top left), \sunset{} (top right), and \fourvertex{} (bottom left) diagrams,
       contributing at $\mathcal{O}(F_\mathrm{NL}^2)$ to the GW power spectrum.
       The bottom right diagram vanishes due to the integration over the azimuthal angles of the
       internal momenta.
    }
   \label{fig:FNL2Feynman}
\end{figure}

\begin{figure}[t]
   \centering
   \includegraphics[width=.48\textwidth]{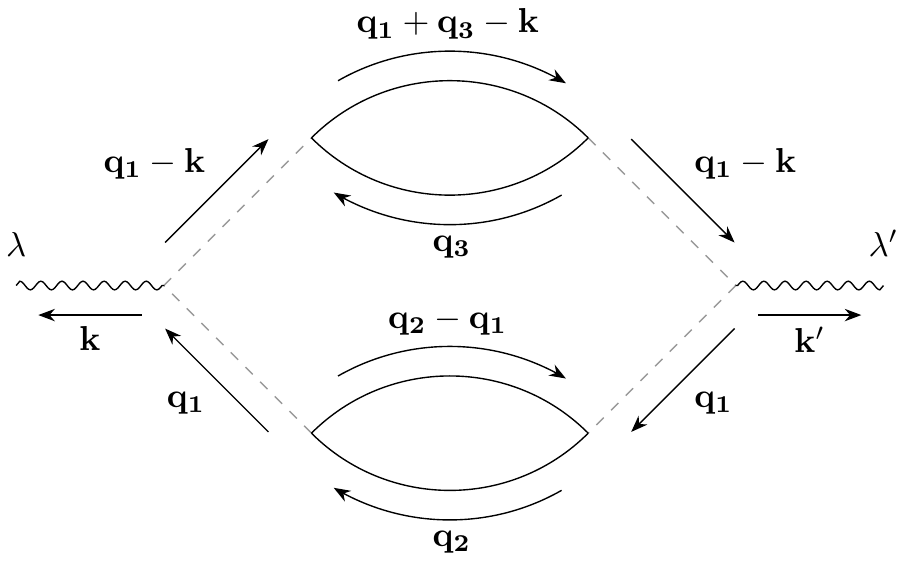}\hfill
   \includegraphics[width=.48\textwidth]{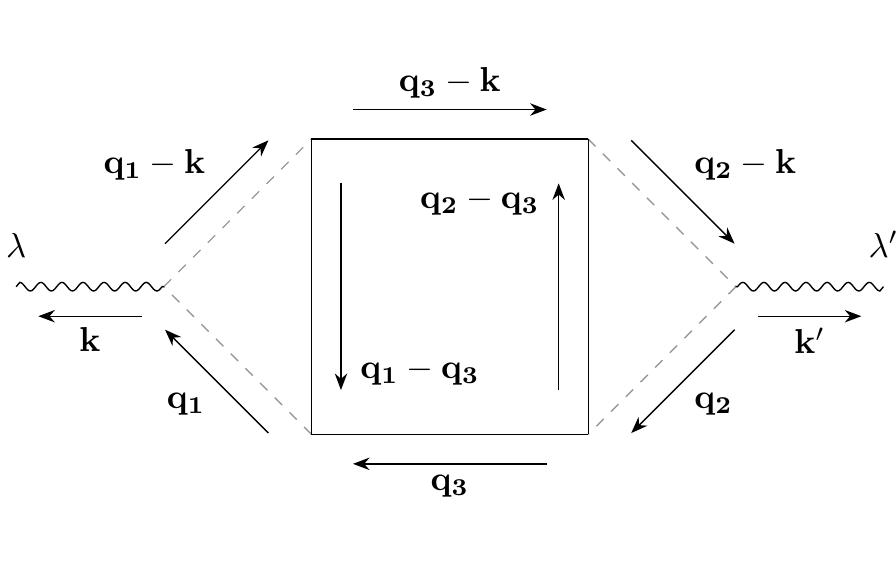}
   \includegraphics[width=.48\textwidth]{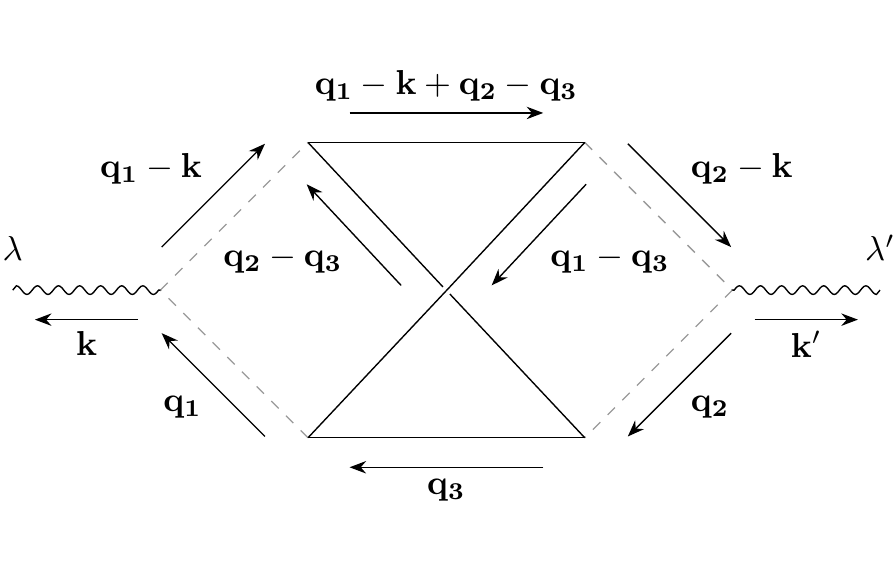}
   \caption{
       The reducible (top left), planar (top right), and nonplanar (bottom) diagrams, contributing
       at $\mathcal{O}(F_\mathrm{NL}^4)$ to the GW power spectrum.
    }
   \label{fig:FNL4Feynman}
\end{figure}

The contribution from the purely Gaussian part of the curvature perturbation is shown in
\cref{fig:FNL0Feynman}.
The contributions due to the non-Gaussianity of the curvature perturbation are shown in
\cref{fig:FNL2Feynman} for those at $\mathcal{O}(F_\mathrm{NL}^2)$ and in
\cref{fig:FNL4Feynman} for those at $\mathcal{O}(F_\mathrm{NL}^4)$.
Contributions at higher order in $F_\mathrm{NL}$ require expanding the stress-energy tensor itself
to higher order in fluctuations and are suppressed by additional powers of the amplitude of the
curvature spectrum.
We have omitted vanishing diagrams in which the solid lines in rule (iii) are connected; these
simply cancel (see \cref{eq:RNG}).


Compared to the diagrams in Ref.~\cite{Unal:2018yaa}, we can identify the hybrid diagram in
\cref{fig:FNL2Feynman} and the disconnected, planar, and nonplanar ones in \cref{fig:FNL4Feynman} by
replacing our dashed lines with solid lines.
However, the so-called ``walnut'' diagram of Ref.~\cite{Unal:2018yaa} could be either of the
remaining diagrams in \cref{fig:FNL2Feynman}.
No rules are provided in Ref.~\cite{Unal:2018yaa} for converting the diagrams to equations.
However, our expression for the \fourvertex{} diagram matches the expression for the ``walnut''
diagram in the supplemental material of Ref.~\cite{Unal:2018yaa}.
(Furthermore, sans transfer functions, the propagator topology of the \sunset{} term more closely
relates to the standard sunset diagram.)

\section{Recasting the integrals}\label{app:recasting-the-integrals}

In this appendix we recast the integrals into a numerically favorable form.
Specifically, we present the polarization-summed, dimensionless gravitational wave power spectrum
\begin{align}
    \Delta_h^2(k)
    &\equiv \frac{k^3}{2 \pi^2}
        \sum_\lambda \mathcal{P}_\lambda(k).
\end{align}
First define the variables
\begin{align}
    u
    &= \frac{\abs{\mathbf{k} - \mathbf{q}}}{k} \\
    v
    &= \frac{q}{k}.
\end{align}
The Jacobian of this transformation is $- k u / v$, so the integration transforms to
\begin{align}
    \int \ud^3 q
    &= k^3
        \int_0^\infty \ud v
        \int_\abs{1 - v}^{1 + v} \ud u \, v u
        \int_0^{2 \pi} \ud \phi.
        \label{eqn:integral-change-vars}
\end{align}
In terms of $u$ and $v$,
\begin{align}
    \sin^2 \theta
    &= \frac{4 v^2 - (1 + v^2 - u^2)^2}{4 v^2}.
\end{align}
Next define
\begin{align}
    s
    &= u - v \\
    t
    &= u + v - 1.
\end{align}
The Jacobian for this transformation is $1/2$, so
\begin{align}
    \int_0^\infty \ud v
    \int_{\abs{1 - v}}^{1 + v} \ud u
    &= \frac{1}{2}
        \int_0^\infty \ud t
        \int_{-1}^1 \ud s.
\end{align}
The integration domain over $s$ and $t$ is rectangular, a requirement of most multidimensional
numerical quadrature routines.
Though these shall be our integration variables, we retain $u$ and $v$ in expressions for notational
convenience.

\Cref{eqn:I-def} takes the form
\begin{align}
    \tilde{I}(v, u, x)
    \equiv k^2 I(q, \abs{\mathbf{k} - \mathbf{q}}, x / k)
    &= \int_{x_0}^x \ud \tilde{x} \,
        k G_{\mathbf{k}}(x / k, \tilde{x} / k) \frac{a(\tilde{x} / k)}{a(x / k)}
        f(v k, u k, \tilde{x} / k).
    \label{eqn:I-to-I-ito-u-v-x}
\end{align}
For further notational convenience, we define
\begin{align}\label{eqn:J-tilde-def}
    \tilde{J}(v, u, x)
    &\equiv v^2 \sin^2 \theta
        \tilde{I}(v, u, x)
    = \frac{4 v^2 - [1 + v^2 - u^2]^2}{4}
        \tilde{I}(v, u, x),
\end{align}
where $\theta$ is the polar angle of $\mathbf{q}$.
Namely, $\tilde{J}(v, u, x)$ is equal to
$\sqrt{2} Q_\lambda(\mathbf{k}, \mathbf{q}) I(q, \abs{\mathbf{k} - \mathbf{q}}, x / k)$ divided by
$\sin 2 \phi$ or $\cos 2 \phi$ for the plus and cross polarizations, respectively.

\subsection{Disconnected contributions}

With the above definitions, the Gaussian contribution \cref{eqn:P-lambda-gaussian} to the
(dimensionless) GW power spectrum (in terms of the dimensionless Gaussian curvature power spectrum
$\Delta_g^2(k)$) is
\begin{align}
    \Delta_h^2(\tau, k)_\mathrm{Gaussian}
    &= 4
        \int_0^\infty \ud t
        \int_{-1}^1 \ud s
        \, u v
        \tilde{J}(v, u, k \tau)^2
        \frac{\Delta_g(v k)}{v^3}
        \frac{\Delta_g(u k)}{u^3}.
\end{align}
Turning to the hybrid term, define
\begin{align}
    u_1
    &= \frac{\abs{\mathbf{k} - \mathbf{q}_1}}{k} \\
    v_1
    &= \frac{q_1}{k} \\
    u_2
    &= \frac{\abs{\mathbf{q}_1 - \mathbf{q}_2}}{q_1} \\
    v_2
    &= \frac{q_2}{q_1}.
\end{align}
Then \cref{eqn:P-lambda-hybrid} yields
\begin{align}
\begin{split}
    \Delta_h^2(\tau, k)_\mathrm{hybrid}
    &= 4 F_\mathrm{NL}^2
        \int_0^\infty \ud t_1
        \int_{-1}^1 \ud s_1
        \int_0^\infty \ud t_2
        \int_{-1}^1 \ud s_2
        \, u_1 v_1^4 u_2 v_2
    \\ &\hphantom{ {}={} 4 F_\mathrm{NL}^2 \int }
        \times
        \tilde{J}(v_1, u_1, k \tau)^2
        \frac{\Delta_g^2(u_1 k)}{u_1^3}
        \frac{\Delta_g^2(v_2 v_1 k)}{(v_2 v_1)^3}
        \frac{\Delta_g^2(u_2 v_1 k)}{(u_2 v_1)^3}.
\end{split}
\end{align}
Further defining
\begin{align}
    v_3
    &= \frac{q_3}{\abs{\mathbf{k} - \mathbf{q}_1}} \\
    u_3
    &= \frac{\abs{ (\mathbf{k} - \mathbf{q}_1) - \mathbf{q}_3 }}{\abs{\mathbf{k} - \mathbf{q}_1}}.
\end{align}
the reducible term, \cref{eqn:P-lambda-reducible}, contributes as
\begin{align}
\begin{split}
    \Delta_h^2(\tau, k)_\mathrm{reducible}
    &= F_\mathrm{NL}^4
        \int_{-1}^{1} \ud s_1
        \int_0^\infty \ud t_1
        \int_{-1}^{1} \ud s_2
        \int_0^\infty \ud t_2
        \int_{-1}^{1} \ud s_3
        \int_0^\infty \ud t_3
        \, u_1^4 v_1^4 u_2 v_2 u_3 v_3
    \\ &\hphantom{{}={} F_\mathrm{NL}^4 \int }
        \times
        \tilde{J}(v_1, u_1, k \tau)^2
        \frac{\Delta_g^2(v_2 v_1 k)}{(v_2 v_1)^3}
        \frac{\Delta_g^2(u_2 v_1 k)}{(u_2 v_1)^3}
        \frac{\Delta_g^2(v_3 u_1 k)}{(v_3 u_1)^3}
        \frac{\Delta_g^2(u_3 u_1 k)}{(u_3 u_1)^3}.
\end{split}
\end{align}

\subsection{Connected diagrams}

For the connected diagrams we define
\begin{align}
    u_i
    &= \frac{\abs{\mathbf{k} - \mathbf{q}_i}}{k} \\
    v_i
    &= \frac{q_i}{k}
\end{align}
for all $i$.
We require the dot products between various $\mathbf{q}_i$,
\begin{align}
\begin{split}
    \frac{\mathbf{q}_i \cdot \mathbf{q}_j}{k^2}
    &= \frac{\cos (\phi_i - \phi_j)}{4}
        \sqrt{
            t_i (t_i + 2) (1 - s_i^2)
            t_j (t_j + 2) (1 - s_j^2)
        }
    \\ &\hphantom{{}={}}
        +
        \frac{1}{4}
        \left[ 1 - s_i (t_i + 1) \right]
        \left[ 1 - s_j (t_j + 1) \right],
        \label{eqn:qi-dot-qj}
\end{split}
\end{align}
as well as between $\mathbf{k}$ and $\mathbf{q}_i$:
\begin{align}\label{eqn:k-dot-qi}
    \frac{\mathbf{k} \cdot \mathbf{q}_i}{k^2}
    = \frac{1}{2} \left[ 1 - s_i (t_i + 1) \right].
\end{align}

Because the integrands only depend on the differences between azimuthal angles, a suitable
coordinate transformation renders one of the azimuthal integrals trivial.
For the \fourvertex{} and \sunset{} terms, \cref{eqn:P-lambda-4-vertex,eqn:P-lambda-sunset},
we integrate over $\varphi = \phi_1 - \phi_2$, leading to
\begin{align}\label{eqn:connected-fnl-squared-ito-s-and-t}
\begin{split}
    \Delta_h^2(\tau, k)_{\mathrm{c}, F_\mathrm{NL}^2}
    &= \frac{4 F_\mathrm{NL}^2}{\pi}
        \prod_{i=1}^2
        \left[
            \int_{-1}^1 \ud s_i
            \int_0^\infty \ud t_i \,
            u_i v_i
            \tilde{J}(u_i, v_i, k \tau)
        \right]
        \int_0^{2 \pi} \ud \varphi \, \cos 2 \varphi
    \\ &\hphantom{{}={} \frac{4 F_\mathrm{NL}^2}{\pi}}
        \times
        \left[
            \frac{\Delta_g^2(v_2 k)}{v_2^3}
            \frac{\Delta_g^2(u_2 k)}{u_2^3}
            \frac{\Delta_g^2(w_a k)}{w_a^3}
            +
            \frac{\Delta_g^2(v_1 k)}{v_1^3}
            \frac{\Delta_g^2(v_2 k)}{v_2^3}
            \frac{\Delta_g^2(w_b k)}{w_b^3}
        \right],
\end{split}
\end{align}
where
\begin{align}
    w_a^2
    &= v_1^2
        + v_2^2
        - 2 \frac{\mathbf{q}_1 \cdot \mathbf{q}_2}{k^2} \\
    w_b^2
    &= 1
        + v_1^2
        + v_2^2
        - 2 \frac{\mathbf{k} \cdot \mathbf{q}_1}{k^2}
        - 2 \frac{\mathbf{k} \cdot \mathbf{q}_2}{k^2}
        + 2 \frac{\mathbf{q}_1 \cdot \mathbf{q}_2}{k^2}.
\end{align}

Finally, for the planar and nonplanar terms, \cref{eqn:P-lambda-planar,eqn:P-lambda-nonplanar}, we
integrate over $\varphi_{12} \equiv \phi_1 - \phi_2$ and $\varphi_{23} \equiv \phi_2 - \phi_3$.
We obtain
\begin{align}
\begin{split}
    \Delta_h^2(\tau, k)_{\mathrm{c}, F_\mathrm{NL}^4}
    &= \frac{F_\mathrm{NL}^4}{2 \pi^2}
        \prod_{i=1}^3
        \left[
            \int_{-1}^1 \ud s_i
            \int_0^\infty \ud t_i \,
            u_i v_i
        \right]
        \prod_{j=1}^2
        \left[
            \tilde{J}(u_j, v_j, k \tau)
        \right]
        \int_{0}^{2 \pi}
        \ud \varphi_{12}
        \ud \varphi_{23} \,
        \cos 2 \varphi_{12}
    \\ &\hphantom{{}={} \frac{F_\mathrm{NL}^4}{2 \pi^2} }
        \times
        \left(
            \frac{\Delta_g^2(v_3 k)}{v_3^3}
            \frac{\Delta_g^2(w_{13} k)}{w_{13}^3}
            \frac{\Delta_g^2(w_{23} k)}{w_{23}^3}
            \left[
                2 \frac{\Delta_g^2(u_3 k)}{u_3^3}
                + \frac{\Delta_g^2(w_{123} k)}{w_{123}^3}
            \right]
        \right),
\end{split}
\end{align}
with the definitions
\begin{align}
    w_{i3}^2
    &= v_i^2
        + v_3^2
        - \frac{2}{k^2} \mathbf{q}_i \cdot \mathbf{q}_3 \\
\begin{split}
    w_{123}^2
    &= 1
        + v_1^2
        + v_2^2
        + v_3^2
        - 2 \frac{\mathbf{k} \cdot \mathbf{q}_1}{k^2}
        - 2 \frac{\mathbf{k} \cdot \mathbf{q}_2}{k^2}
        + 2 \frac{\mathbf{k} \cdot \mathbf{q}_3}{k^2}
    \\ &\hphantom{ {}={} }
        - 2 \frac{\mathbf{q}_3 \cdot \mathbf{q}_1}{k^2}
        - 2 \frac{\mathbf{q}_3 \cdot \mathbf{q}_2}{k^2}
        + 2 \frac{\mathbf{q}_1 \cdot \mathbf{q}_2}{k^2}.
\end{split}
\end{align}

We implement the integrals numerically with \textsf{vegas+}~\cite{Lepage:2020tgj}, performing each
for at least 200 external momenta $k$ and with a sufficiently large number of evaluations to achieve
a relative precision of $10^{-3}$.
For the higher dimensional integrals (the reducible and all connected terms), we find using a
relatively short MCMC sample (implemented with \textsf{emcee}~\cite{ForemanMackey:2012ig}) as a
preconditioner to be an efficient means of generating an optimal \textsf{vegas} map (see the
discussion in Ref.~\cite{Lepage:2020tgj}).

\section{Monochromatic spectrum: infrared limit}\label{sec:monochromatic-analytic}

We now sketch a derivation of the IR scaling of the induced GW spectrum for the monochromatic
case, considering the \fourvertex{} term [\cref{eqn:P-lambda-4-vertex}] as an example.
Starting from \cref{eqn:connected-fnl-squared-ito-s-and-t}, changing integration variables from
$\varphi$ to $c \equiv \cos \varphi$, and substituting \cref{eqn:P-g-monochromatic},
\begin{align}
\begin{split}
    \frac{
        \Delta_h^2(k, \tau)_{\mathrm{\fourvertex{}}}
    }{
        F_\mathrm{NL}^2 \mathcal{A}_\mathcal{R}^3
    }
    &= \frac{4}{\pi}
        \prod_{i=1}^2
        \left[
            \int_{-1}^1 \ud s_i
            \int_0^\infty \ud t_i \,
            u_i v_i
            \tilde{J}(u_i, v_i, k \tau)
        \right]
        2 \int_{-1}^{1} \ud c \,
        \frac{2 c^2 - 1}{\sqrt{1 - c^2}}
    \\ &\hphantom{{}={} \frac{4}{\pi}}
        \times
            \frac{\delta(v_1 \tilde{k} - 1)}{v_1^3}
            \frac{\delta(u_1 \tilde{k} - 1)}{u_1^3}
            \frac{\delta(w_a \tilde{k} - 1)}{w_a^3}.
\end{split}
\end{align}
Recall that $\tilde{k} = k / k_\star$.
Integrating the Dirac delta functions over $s_1$, $t_1$, and $c$ sets
\begin{subequations}
\label{eqn:monochromatic-C-values-for-integration-vars}
\begin{align}
    s_1
    &= 0 \\
    t_1
    &= 2 / \tilde{k} - 1 \\
    c
    &= \frac{- 1 + u_2^2 + v_2^2}{4 v_1 v_2 \sin \theta_1 \sin \theta_2},
    \label{eqn:monochromatic-C-value-for-c}
\end{align}
\end{subequations}
and introduces a Jacobian factor of $2 w_a / \tilde{k}^3 v_1 v_2 \sin \theta_1 \sin \theta_2$.
Here $\theta_i$ is the azimuthal angle of $\mathbf{q}_i$, for which
$\sin^2 \theta_i = t_i (t_i + 2) (1 - s_i^2) / 4 v_i^2$.
Note that \cref{eqn:monochromatic-C-values-for-integration-vars} sets
$v_1 \sin \theta_1 = \sqrt{1 - \tilde{k}^2 / 4} / \tilde{k}$.
The leading-order, IR behavior of the product of the (late-time, oscillation-averaged)
transfer functions, \cref{eqn:late-time-transfer-function-product}, resides in the
$\tilde{I}_A \tilde{I}_B$ factors.
After substituting \cref{eqn:J-tilde-def} for $\tilde{J}$ and evaluating $\tilde{I}_A(u_1, v_1)
\tilde{I}_B(u_1, v_1)$ at $u_1 = v_1 = 1 / \tilde{k}$,
\begin{align}
\begin{split}
    \frac{
        \Delta_h^2(k, \tau)_{\mathrm{\fourvertex{}}}
    }{
        F_\mathrm{NL}^2 \mathcal{A}_\mathcal{R}^3
    }
    &\approx \frac{12 \tilde{k}^4}{2 \pi (k \tau)^2}
        \ln\left( \frac{4}{3 \tilde{k}^2} \right)
        \Theta(2 - \tilde{k})
    \\ &\hphantom{
            {}={}
        }
        \times \int_{-1}^1 \ud s_2
        \int_0^\infty \ud t_2 \,
        \frac{2 (2 c^2 - 1) \Theta(1 - \abs{c})}{\sqrt{1 - c^2}}
        \tilde{I}_A(u_2, v_2)
        \tilde{I}_B(u_2, v_2)
        u_2 v_2^2 \sin \theta_2,
\end{split}
\end{align}
where $c$ is implicitly set by \cref{eqn:monochromatic-C-value-for-c}.

The $\tilde{k}$-dependence of the remaining integral over $s_2$ and $t_2$ arises solely from $c$'s
own $\tilde{k}$-dependence, affecting both the integrand and the bounds of integration via the
Heaviside function $\Theta(1 - \abs{c})$.
Inspecting the form of $c$ in \cref{eqn:monochromatic-C-value-for-c} (in terms of $s_2$ and $t_2$)
shows that this Heaviside factor cuts off the integrals at $t_2 \sim 4 / \tilde{k}$.
Furthermore, the integrand grows with $t_2$, so the integral is dominated by this upper limit.
The integrand depends weakly on $s_2$ and one can show that regardless of $s_2$'s value, to leading
order in $\tilde{k}$ and $1/t_2$ the $t_2$-dependence of the integrand is $\sim t_2 \ln t_2$.
As a result, the integral itself contributes a factor $\sim \tilde{k}^{-2} \ln \tilde{k}$ and so, to
leading order in $\tilde{k}$,
\begin{align}
    \Omega_{\mathrm{GW},0}
    \sim \tilde{k}^2 \ln^2 \tilde{k}.
\end{align}

\bibliography{BibAuto}
\bibliographystyle{JHEP}

\end{document}